\documentclass[preprint,sort&compress,12pt]{elsarticle}

\usepackage{amssymb}
\usepackage{amsthm}
\usepackage{amsmath}
\usepackage{mathtools}
\usepackage{eucal}
\usepackage{graphicx}
\usepackage[font=normalsize,labelfont=bf]{caption}
\usepackage[font=normalsize]{subfig}
\usepackage{algorithm}
\usepackage{algpseudocode}
\usepackage{array}
\usepackage{multirow}
\usepackage{enumerate}
\usepackage{diagbox} 
\usepackage{soul}
\usepackage[dvipsnames]{xcolor}
\definecolor{color0}{rgb}{0.12156862745098,0.466666666666667,0.705882352941177}
\definecolor{color1}{rgb}{0.647058823529412,0.164705882352941,0.164705882352941}
\definecolor{color2}{rgb}{1,0.498039215686275,0.0549019607843137}
\definecolor{color3}{rgb}{0.172549019607843,0.627450980392157,0.172549019607843}
\definecolor{color4}{rgb}{0.83921568627451,0.152941176470588,0.156862745098039}
\usepackage{tikz}
\usepackage[utf8]{inputenc}
\usepackage{pgfplots}
\pgfplotsset{compat=newest}
\usepgfplotslibrary{groupplots}
\usepgfplotslibrary{dateplot}
\usepackage{tikzscale}
\usetikzlibrary{arrows,automata}
\setcitestyle{square}
\usepackage{lineno}
\usepackage{fullpage}
\usepackage{xcolor}
\usepackage[colorinlistoftodos]{todonotes}
\usepackage{bbm}
\usepackage{listings}
\usepackage[colorlinks=true]{hyperref}
\usepackage{url}

\graphicspath{ {./figs/} }
\biboptions{comma,round}

\graphicspath{ {./figs/} }

\journal{Elsevier}

\DeclareUnicodeCharacter{2212}{-}

\usepackage{changes}
\definechangesauthor[name={Reviewer 1}, color = blue]{R1}
\definechangesauthor[name={Reviewer 2}, color = red]{R2}
\definechangesauthor[name={Reviewer 1 \& 2}, color = olive]{All}
\definechangesauthor[name={Reviewer 3}, color = brown]{R3}
\definechangesauthor[name={Reviewer 4}, color = green]{R4}

\begin{document}
\begin{frontmatter}

 \title{PhyGeoNet: Physics-Informed Geometry-Adaptive Convolutional Neural Networks for Solving {Parameterized Steady-State} PDEs on Irregular Domain}



\author[ndAME,ndCICS]{Han Gao}
\author[ndAME,ndCICS]{Luning Sun}
\author[ndAME,ndCICS]{Jian-Xun Wang\corref{corxh}}

\address[ndAME]{Department of Aerospace and Mechanical Engineering, University of Notre Dame, Notre Dame, IN}
\address[ndCICS]{Center for Informatics and Computational Science, University of Notre Dame, Notre Dame, IN}
\cortext[corxh]{Corresponding author. Tel: +1 540 3156512}
\ead{jwang33@nd.edu}

\begin{abstract}
Recently, the advent of deep learning has spurred interest in the development of physics-informed neural networks (PINN) for efficiently solving partial differential equations (PDEs), particularly in a parametric setting. Among all different classes of deep neural networks, the convolutional neural network (CNN) has attracted increasing attention in the scientific machine learning community, since the parameter-sharing feature in CNN enables efficient learning for problems with large-scale spatiotemporal fields. However, one of the biggest challenges is that CNN only can handle regular geometries with image-like format (i.e., rectangular domains with uniform grids). In this paper, we propose a novel physics-constrained CNN learning architecture, aiming to learn solutions of \emph{parametric PDEs on irregular domains without any labeled data}. In order to leverage powerful classic CNN backbones, elliptic coordinate mapping is introduced to enable coordinate transforms between the irregular physical domain and regular reference domain. The proposed method has been assessed by solving a number of {steady-state} PDEs on irregular domains, including {heat equations, Navier-Stokes equations, and Poisson equations with parameterized boundary conditions, varying geometries, and spatially-varying source fields}. Moreover, the proposed method has also been compared against the state-of-the-art PINN with fully-connected neural network (FC-NN) formulation. The numerical results demonstrate the effectiveness of the proposed approach and exhibit notable superiority over the FC-NN based PINN in terms of efficiency and accuracy.

\end{abstract}

\begin{keyword}
  Physics-informed neural networks \sep Label-free \sep Surrogate modeling \sep Physics-constrained deep learning \sep Partial differential equations \sep Navier-Stokes
\end{keyword}
\end{frontmatter}


\section{Introduction}
\label{sec:intro}
Physical phenomena in science and engineering are usually modeled by partial differential equations (PDEs), whose states live in infinite-dimensional spaces. Due to the lack of analytical solutions in most cases, their finite-dimensional approximations are resorted to based on traditional numerical approaches, e.g., finite difference (FD), finite volume (FV), and finite element (FE) methods, which have been developed and advanced over the past several decades. Nonetheless, the traditional numerical solvers often require significant computational efforts particularly for complex systems with multiscale/multiphysics features and might not even be feasible in real-time or many-query applications, e.g., optimization, inverse problem, and uncertainty quantification (UQ), where a large number of repeated simulations are required. Solving PDE systems with an optimal balance between accuracy and efficiency still remains a long-standing challenge. Recently, deep neural networks (DNNs) have been demonstrated to be promising for solving PDEs or metamodeling of PDE-based systems~\cite{sirignano2018dgm,raissi2019physics,zhu2019physics,sun2020surrogate,brunton2020machine,zhang2020physics}. The advantages of using DNNs to approximate solutions of PDEs can be summarized as follows: (i) DNNs are capable to express any strong nonlinear relationships, mathematically supported by universal approximation theorems~\cite{hornik1989multilayer,cybenko1989approximation}; (ii) forward evaluations of trained DNNs are extremely efficient, which is a desirable feature for real-time or many-query applications; (ii) DNN models are analytically differentiable and thus derivative information can be easily extracted via automatic differentiation for optimization and control problems. 

\subsection{Physics-informed fully-connected neural networks}
Generally, training of a DNN model requires a vast amount of labeled data, which are often unavailable in many scientific machine learning (SciML) applications~\cite{wang2017physics,wang2016physics,roscher2020explainable,peng2020multiscale}. However, when the governing PDEs are known, their solutions can be learned in a physics-constrained fashion with less data~\cite{raissi2019physics,zhang2019physics,sun2020physics} or even without any data~\cite{zhu2019physics,geneva_modeling_2020,sun2020surrogate}. Namely, physics-informed loss functions are constructed based on PDE residuals and the DNN is trained by minimizing the violation of physical laws. The idea of using neural networks to solve differential equations has been proposed decades ago~\cite{lee1990neural,lagaris1998artificial,lagaris2000neural} and recently experienced a resurgence of interest due to remarkable advances in deep learning~\cite{raissi2019physics,michoski2020solving}. Raissi et al.~\cite{raissi2019physics} developed physics-informed neural networks (PINN) using modern deep learning techniques, where the PDE residual is incorporated into the loss function of fully-connected neural networks (FC-NN) as a regularizer, enabling training in ``small data" regimes. This weakly-supervised method has been applied to solve various PDEs with limited training data in many scientific problems, including subsurface flows~\cite{wang2020deep}, vortex-induced vibrations~\cite{raissi2019deepviv}, turbulent flows~\cite{raissi2019deep,jin2020nsfnets}, cardiovascular systems~\cite{sun2020physics,raissi2020hidden,kissas2020machine,sahli2020physics}, metamaterial design~\cite{fang2019deep,liu2019multi,chen2019physics}, geostatistical modeling~\cite{zheng2019physics}, and others~\cite{chen2019learning,snaiki2019knowledge,mao2020physics}. This idea has also been extended to utilize multi-fidelity datasets~\cite{meng2020composite} and solve system identification problems~\cite{tartakovsky2018learning,berg2019data,lu2019deeponet}. In order to handle sparse, noisy data and quantify aleatoric uncertainty arising from measurement noise, a Bayesian formulation of the physics-constrained learning was proposed by Sun and Wang~\cite{sun2020physics} using variational inference (VI), and a more comprehensive study on Bayesian PINN was conducted by Yang et al.~\cite{yang2020b}, where both the VI and Hamiltonian Monte Carlo formulations are compared. The total uncertainty quantification (UQ) in PINN was also investigated based on adversarial inference~\cite{yang2019adversarial} and arbitrary polynomial chaos in conjunction with dropout~\cite{zhang2019quantifying}. Although a moderate amount of training data are needed in the aforementioned works, the requirement of data can be entirely avoided if initial/boundary conditions (IC/BC) are properly enforced. Namely, the PDEs with specified IC/BCs can be solved deterministically within a deep learning framework. The effectiveness of the data-free FC-NN based PDE solution algorithm has been demonstrated on a number of canonical PDEs~\cite{lu2019deepxde,berg2018unified} and stochastic PDEs~\cite{sirignano2018dgm,yang2020physics,han2018solving}. To improve the learning performance, researchers have recently explored several different directions, e.g., strong/variational formulations of PDE residuals~\cite{zang2020weak,samaniego2020energy,khodayi2019varnet,kharazmi2019variational}, distributed learning using domain decomposition~\cite{li2019d3m,kharazmi2020hp,dwivedi2019physics}, and convergence analysis in network training/optimizations~\cite{zhang2020machine,wang2020understanding,lu2019deepxde}.

\subsection{Physics-informed convolutional neural networks}
Although physics-informed FC-NN for deterministically solving PDEs has been extensively investigated, the solution predictions for parametric PDEs in parameterized spaces (e.g., IC/BCs, geometry, and equation parameters) are less explored. The latter is critical for developing efficient metamodels (i.e., surrogate models). Only a very few existing studies aimed to build surrogate models using physics-informed FC-NN. Nabian and Meidani~\cite{nabian2019deep} and Karumuri et al.~\cite{Karumuri2020Sim} applied the physics-informed FC-NN for efficient uncertainty propagation in steady heat equations. Sun et al.~\cite{sun2020surrogate} developed a physics-constrained DNN surrogate for fluid flows with varying geometries and fluid properties, which was the first attempt of using FC-NNs to learn the solutions of parametric Navier-Stokes equations without relying on any labeled data. Despite some success, the FC-NN formulation suffers from scalability issues. This is because the training cost of FC-NNs will significantly increase for complex problems especially with parametric variations since the PDE residuals need to be evaluated on massive amounts of collocation points in high-dimensional input spaces. To enable efficient learning of large-scale spatiotemporal solution fields, the convolutional neural network (CNN) structure has attracted increasing attention in the SciML community. Compared to FC-NNs, CNNs usually need orders of magnitude fewer parameters because of parameter sharing via filter-based convolution operations, which is thus well-positioned for large-scale and high-dimensional problems \cite{zhu2019physics,rao2020three}. 

Most recently, researchers started to blend physics with CNN-based learning, such as imposing physical constraints into CNNs~\cite{kim2019deep,sharma2018weakly,fukui2019physics,subramaniam2020turbulence,mohan2020embedding,wu2020enforcing,yao2019fea}, development of physics-informed CNN for surrogate modeling~\cite{he2019mgnet,zhu2019physics,geneva_modeling_2020,joshigenerative,zhang2020physics}, and discovering underlying governing equations from observed data within CNN-based architectures~\cite{long2017pde,long2019pde,singh2020time,rackauckas2020universal}. Kim et al.~\cite{kim2019deep} presented a generative CNN model to synthesize fluid simulations with the strictly imposed divergence-free condition using stream functions, and a similar idea was investigated by Mohan et al.~\cite{mohan2020embedding} for coarse-graining of 3D turbulent flows, {where boundary conditions were also strictly enforced on a regular domain.} The physical constraints can also be imposed in a soft manner as penalty terms by modifying loss functions, and this idea has been exploited to enable weakly-supervised learning (i.e., using less labeled data)~\cite{sharma2018weakly} or physically-correct synthetic turbulence generation and turbulence enrichment~\cite{wu2020enforcing,subramaniam2020turbulence}. In terms of surrogate modeling, Zhu et al.~\cite{zhu2019physics} devised a physics-constrained convolutional encoder-decoder structure with a conditional-based generative model to learn solutions of high-dimensional elliptic PDEs without using any labeled data, and this approach has recently been extended for learning dynamical PDEs with parameterized initial conditions~\cite{geneva_modeling_2020}. In a similar vein, Joshi et al.~\cite{joshigenerative} generated solutions of parametric Burgers equations by CNNs using adversarial learning. As for equation discovery, Long et al.~\cite{long2017pde,long2019pde} utilized deep CNNs in combination with symbolic networks to discover PDEs from spatiotemporal data by interpreting the learned filters. Rackauckas et al.~\cite{rackauckas2020universal} developed universal differential equations (UDEs) that leverage CNN to discover unknown equations from data. Singh et al.~\cite{singh2020time} presented a low-weight interpretable convolutional encoder-decoder network to capture the invariant structure of observation data for various PDE systems.

\subsection{Scope and contributions of present work}
Despite showing great promise, all the existing studies on physics-informed CNNs are only able to deal with problems defined on \emph{regular (rectangular) domains with uniform grids}, which largely limits their applications to general scientific problems, where geometries are often \emph{complex and irregular}. The underlying reason is that classic CNNs and their convolution operations are originally designed for processing natural images, which are described as functions in Euclidean space, sampled on a uniform mesh. However, the coordinate frames for problems with irregular domains have non-Euclidean structures, where the shift invariance that justifies the use of classic convolutional filters is no longer valid. Particularly in physics-constrained learning, the derivative terms in PDE-informed loss are computed based on finite difference through convolution operations, which only works in image-like rectangular domains. To handle data with non-Euclidean structures, in the Artificial Intelligence (AI) community, there recently has been growing interest in geometric deep learning, which is an {umbrella} term for emerging techniques aiming to adapt CNN to non-Euclidean space. Graph theory, spectral transformation, and manifold embedding, etc., have been utilized to reformulate the non-Euclidean convolution operations. Nonetheless, many of these newly proposed geometric CNNs have difficulties generalizing across different topologies, and moreover, it is not clear how to construct PDE-based loss function and impose boundary conditions for physics-constrained learning. 

In this work, we propose a \emph{novel} physics-constrained deep learning method, named as the physics-informed geometry-adaptive convolutional neural network (PhyGeoNet), \emph{aiming to learn solutions of parametric PDEs on irregular domains without using any labeled data} (no simulations data are needed for training). Specifically, we use elliptic coordinate transformations to adapt CNN-based learning for problems with irregular geometries. Unlike the graph/geodesic CNNs, where problem-specific convolutional filters have to be designed, our proposed method can directly leverage powerful state-of-the-art uniform Cartesian-grid-based CNN architectures. Moreover, PDEs on the reference domain will be used to construct physics-informed loss function via finite-difference convolution kernels and boundary conditions are strictly encoded into networks based on padding operations. The \emph{novel contributions} of this paper are as follows: (a) we propose a physics-informed CNN architecture, enabling data-free learning for parametric PDEs with irregular geometries; (b) encode boundary conditions into the CNN architecture in a hard manner~{(i.e., BCs are strictly enforced~\cite{sun2020surrogate,mohan2020embedding} for irregular geometries)}; (c) demonstrate the effectiveness of the proposed method on parametric heat equations and Navier-Stokes equations; (d) compare the proposed method with physics-informed FC-NNs (i.e., PINN) in terms of accuracy and efficiency. Moreover, to the best of authors’ knowledge, this is the first attempt of using CNN to learn parametric Navier–Stokes equations on complex geometries without relying on any labeled data for training. The rest of the paper is organized as follows. The framework of proposed PhyGeoNet is introduced in Section~\ref{sec:meth}. Numerical results on heat equations and Navier-Stokes equations in both non-parametric and parametric settings are presented in Section~\ref{sec:result}. The performance comparison between PhyGeoNet and PINN is discussed in Section~\ref{sec:discussion}. Finally, Section~\ref{sec:conclusion} concludes the paper.


%
%

\section{Methodology\label{sec:meth}}
Physics-constrained learning is formulated by constructing a PDE-based loss function, where PDE residuals with the neural network (NN)-approximated solutions are computed. In CNN architecture, the derivative terms are obtained by finite difference using spatially-invariant convolution operations, which fails on irregular domains with non-uniform grids. In this section, we first provide a mathematical overview of physics-constrained learning using classic CNN and discuss its limitations for complex geometries. Then, we introduce a novel learning framework of the geometry-adaptive CNN, PhyGeoNet, where elliptic coordinate transformation is encoded to handle non-uniform grids and irregular geometries. The mathematical concept, network architecture, and implementation details are discussed.

\subsection{Physics-constrained learning with classic convolutional neural network}
\subsubsection{Learning for PDE solutions on uniform grids without labels}
The goal of this work is to learn the steady-state solutions of PDEs in a parametric setting. Consider a system of parametric {steady} PDEs, 
\begin{equation}
	\label{eqn:generalSteadyPDE}
	\begin{split}
		\mathcal{F}(\mathbf{u},\nabla\mathbf{u}, \nabla^2 \mathbf{u}, \cdots; \boldsymbol{\mu})=0,\;\;\text{in }\Omega_{p}, \\
		\mathcal{B}(\mathbf{u},\nabla\mathbf{u}, \nabla^2 \mathbf{u}, \cdots; \boldsymbol{\mu})=0,\;\;\text{on }\partial\Omega_{p},
	\end{split}
\end{equation}
where $\mathcal{F}(\cdot)$ represents PDE operators, defined on the physical domain $\Omega_{p}$ and parameterized by $\boldsymbol{\mu}(\mathbf{x})$; The solution variables of the PDE system are denoted by $\mathbf{u} = \mathbf{u}(\mathbf{x})$, which is a scalar or vector function of spatial coordinates $\mathbf{x}\in\Omega_p$; $\nabla$ is the gradient operator with respect to $\mathbf{x}$; $\mathcal{B}(\cdot)$ represents functions for boundary conditions (BCs), which are enforced on the boundary $\partial\Omega_{p}$ of the physical domain. When a set of parameters $\boldsymbol{\mu}$ is given, the PDE system (Eq.~\ref{eqn:generalSteadyPDE}) can be solved numerically using FD, FV, or FE methods, which is usually time-consuming. However, such a process has to be performed each time $\boldsymbol{\mu}$ is changed, posing great challenges on UQ/optimization applications. To enable rapid forward propagation of coordinates/parameters [$\mathbf{x}, \boldsymbol{\mu}$] to state variables $\mathbf{u}(\mathbf{x}; \boldsymbol{\mu})$, neural networks are introduced to learn the parameter-to-state map. The learning process can be done completely offline and the online inference for new parameters will be very efficient via the trained network. In most existing works, FC-NN architectures were employed to learn the mapping in a pointwise fashion, which, however, introduces considerable training burden for large-scale problems. Unlike FC-NNs, CNNs enable image-based end-to-end learning and can directly generate solution fields over the entire domain instead of pointwise outputs. Hence, they often have much less training costs and better predictive accuracy~\cite{zhu2019physics}. Specifically, the discrete solution field $\mathbf{u}(\boldsymbol{\chi}, \boldsymbol{\mu}(\boldsymbol{\chi}))$ on a \emph{rectangular domain} can be approximated by a CNN model,
\begin{equation}
	\label{eqn:cnn-approx}
	\mathbf{u}\Big(\boldsymbol{\chi}, \boldsymbol{\mu}(\boldsymbol{\chi})\Big) \approx  \mathbf{u}^{cnn}\Big(\boldsymbol{\chi}, \boldsymbol{\mu}(\boldsymbol{\chi}); \Gamma\Big),
\end{equation}
where $\boldsymbol{\chi} = \{\mathbf{x}_1, \cdots, \mathbf{x}_{n_g}\}$ represents a set of $n_g$ fixed grid points uniformly spaced (like pixels/voxels of images); $\Gamma = \{ \gamma^{l} \} _{l=1}^{n_l}$ is a bank of trainable filters for convolution operations. The input layer of the CNN model includes discrete spatial coordinate $\boldsymbol{\chi}$ and parameter fields $\boldsymbol{\mu}(\boldsymbol{\chi})$, which can be formulated as multiple image channels. To obtain the output solutions, multiple convolutional (conv) layers are acted on the input channels by applying the bank of filters $\Gamma$ and pointwise nonlinearity $\phi(\cdot)$ (i.e., activation functions). For instance, the output (i.e., feature map) $g^l(\mathbf{x})$ of the $l^{\text{th}}$ conv layer can be expressed as,
\begin{equation}
	g^l(\mathbf{x}) = \phi\Big((g^{l}\odot \gamma^{l})(\mathbf{x})\Big), \  \mathbf{x} \in \boldsymbol{\chi}^l,
\end{equation}
where $\odot$ denotes convolution operation,
\begin{equation}
	(g\odot \gamma)(\mathbf{x}) = \int_{\boldsymbol{\chi}}{g(\mathbf{x} - \mathbf{x}')\gamma(\mathbf{x}')d\mathbf{x}'}.
\end{equation}
By convention, one can train the CNN by minimizing the cost function on the training set $\{\boldsymbol{\mu}_i,  \mathbf{u}^{d}_i\}_{i=1}^{n_d}$,
\begin{equation}
	\label{eqn:datatraining}
	\min_{\Gamma} \sum_{i=1}^{n_d} \underbrace{\left\| \mathbf{u}^{cnn}\big(\boldsymbol{\chi}, \boldsymbol{\mu}_i; \Gamma\big) - \mathbf{u}^{d}_i\big(\boldsymbol{\chi}\big) \right\|_{\Omega_{p}}}_{\text{data-based loss:}\ \mathcal{L}_{data}},
\end{equation}
where $\| \cdot \|_{\Omega}$ is $L_2$ norm over spatial domain ${\Omega}_{p}$. Such purely data-based learning requires enormous ($n_d$) labeled data $\mathbf{u}_i^d, i = 1, \cdots, n_d$, which are usually too expensive to obtain from either numerical simulations or experiments. As an alternative, the training can be formulated as a constrained optimization using governing equations and corresponding BCs,
\begin{equation}
	\label{eqn:pdetraining}
	\begin{split}
		&\min_{\Gamma} \sum_{i=1}^{n_d}
		\underbrace{\left\|
			\mathcal{F}\Big(\mathbf{u}^{cnn}(\boldsymbol{\chi}, \boldsymbol{\mu}_i; \Gamma),
			\nabla\mathbf{u}^{cnn}(\boldsymbol{\chi}, \boldsymbol{\mu}_i; \Gamma), 
			\nabla^2\mathbf{u}^{cnn}(\boldsymbol{\chi}, \boldsymbol{\mu}_i; \Gamma), \cdots; \boldsymbol{\mu}_i\Big) \right\|_{\Omega_{p}}}_{\text{equation-based loss:}\ \mathcal{L}_{pde}},\\
		&s.t.\;\; \mathcal{B}\Big(\mathbf{u}^{cnn}(\boldsymbol{\chi}, \boldsymbol{\mu}_i; \Gamma),
		\nabla\mathbf{u}^{cnn}(\boldsymbol{\chi}, \boldsymbol{\mu}_i; \Gamma), 
		\nabla^2\mathbf{u}^{cnn}(\boldsymbol{\chi}, \boldsymbol{\mu}_i; \Gamma), \cdots;
		\boldsymbol{\mu}_i\Big)=0,\;\;\text{on }\partial\Omega_{p}.
	\end{split}
\end{equation}
The spatial derivative terms (i.e., $\nabla\mathbf{u}^{cnn}, \nabla^2\mathbf{u}^{cnn}$) of the CNN ansatz are numerically approximated via convolution operation with predefined non-trainable filters, which are interpreted as discretized differential operators in the finite difference form~\cite{long2017pde,long2019pde,zhu2019physics}. Constraints of BCs can be either treated as penalty terms (soft BC enforcement)~\cite{raissi2019physics,zhu2019physics} or strictly encoded into the network architecture (i.e., hard BC enforcement)~\cite{sun2020surrogate}. The optimization problem defined in Eq.~\ref{eqn:pdetraining} is solved based on stochastic gradient descent (SGD), e.g., Adam algorithm~\cite{kingma2014adam}. In the optimization, only residuals of the PDEs are evaluated without the need of solving them. Note that if a small amount of labeled data are available, the equation-based loss ($\mathcal{L}_{pde}$ in Eq.~\ref{eqn:pdetraining}) can be combined with the data-based loss ($\mathcal{L}_{data}$ in Eq.~\ref{eqn:datatraining}), and then the combined loss function (A.K.A. physics-informed loss) will be minimized, as weakly-supervised learning. In this work, we only focus on purely PDE-driven learning without using any labeled data. Namely, neural networks are used to solve the PDEs in a parametric setting.

\subsubsection{Limitations of classic CNN on irregular geometries with nonuniform grids}
\begin{figure}[H]
	\centering
	\subfloat[Non-uniform grids]
	{\includegraphics[width=0.38\textwidth]{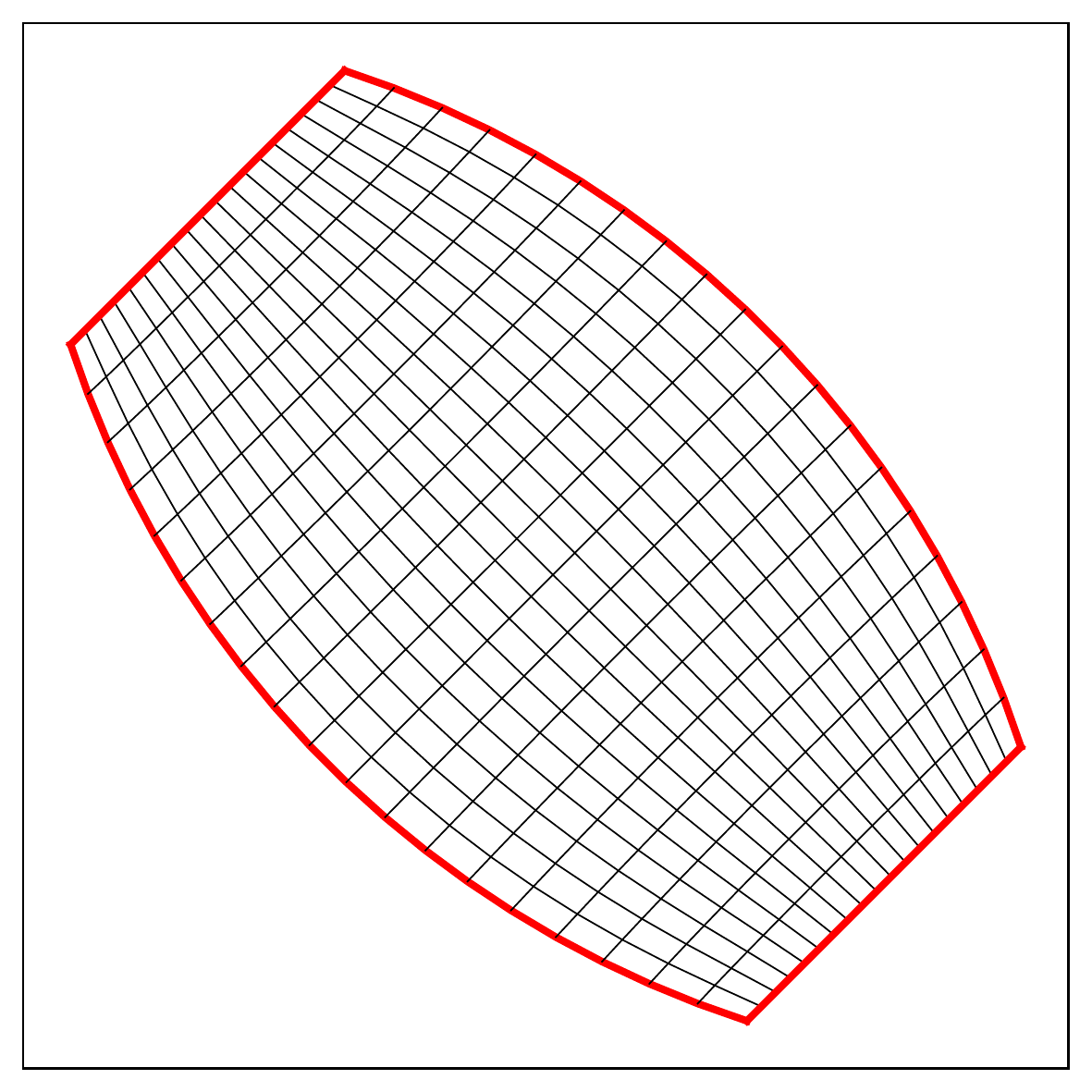}}
	\subfloat[Rasterization]
	{\includegraphics[width=0.38\textwidth]{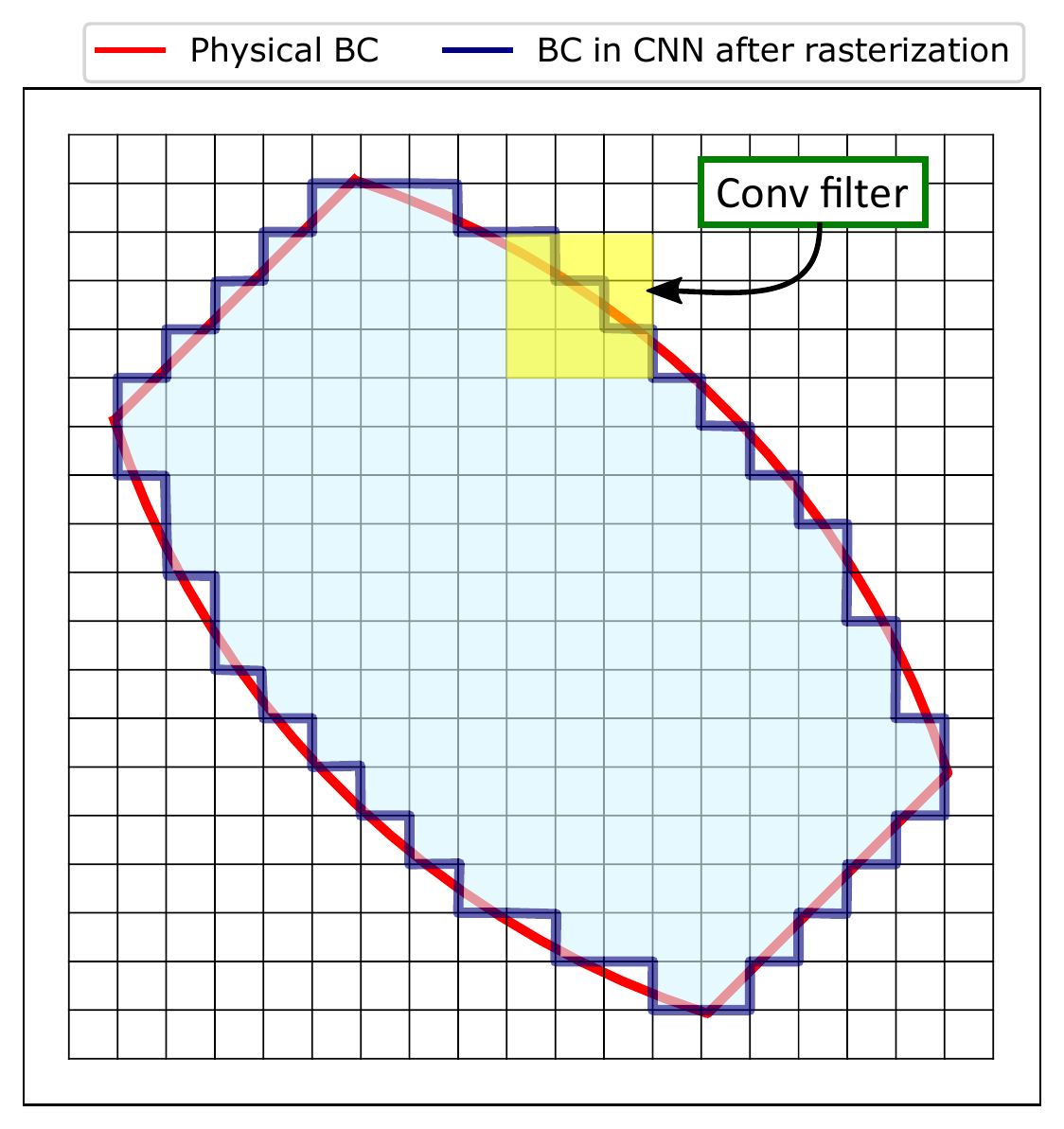}}
	\caption{Limitations of the classic CNN in solving PDEs on an irregular domain.}
	\label{fig:TraditionCNN}
\end{figure}

The classic CNNs had originally been developed for image recognition and processing, where the convolution filters are designed for images or image-like data. Namely, the discrete coordinates $\boldsymbol{\chi}$ in Eq.~\ref{eqn:cnn-approx} have to be a set of uniform grid points within a rectangular domain. However, in scientific computing and physical modeling applications, the geometries are usually irregular with non-uniform grids (e.g., boundary-fitted mesh in Fig.~\ref{fig:TraditionCNN}a). 
In such cases, classic CNN techniques are not directly applicable since the Euclidean-distance-based convolution filters are no longer invariant on non-uniform meshes. Although new convolution filters can be defined for a given boundary-fitted coordinate, how to generalize them across different domain shapes (e.g., in geometric parameterization) and how to construct differential operators for physics-based loss function are still challenging. To enable direct use of the classic CNN backbone, ``rasterization" is often performed to preprocess the data in data-driven scenarios, converting irregular shapes into uniform-grid based 2D/3D images, where pixels/voxels are labeled by binaries~\cite{han2018flownet,han2019tsr,han2019flow} or Signed Distance Function (SDF) to represent geometries~\cite{bhatnagar2019prediction, guo2016convolutional}, as shown in Fig.~\ref{fig:TraditionCNN}b. Nonetheless, the binary or SDF based geometry representations fail to operate in physics-constrained learning due to the following issues and drawbacks:
\begin{enumerate}
	\item  After rasterization, the boundary shape becomes stepped and zigzag (Fig.~\ref{fig:TraditionCNN}), introducing big aliasing errors under low/moderate image resolution. Especially, when it comes to boundary-value problems, even a minor misrepresentation of the boundary may lead to large errors in the learned solution field. 
	\item Using binary/SDF representations, it is difficult or even impossible to impose PDE boundary conditions ($\mathcal{B}(\mathbf{u}^{cnn}) = 0$) and thus physics-constrained learning would fail, particularly when labeled data are scarce or even absent. This is because the PDE solutions are uniquely determined by given BCs, and without properly prescribed BCs the optimization problem is ill-posed.
	\item When the shape of the physical domain (blue region in Fig.~\ref{fig:TraditionCNN}b) is far from its corresponding rectangular envelope, the artifacts introduced from the background region (blank region in Fig.~\ref{fig:TraditionCNN}b) may complicate the training process and make the optimization easier to trap in bad local minima. For instance, in binary-based representation, the derivatives computed in the background region are always zeros and thus the PDEs are exactly satisfied, which clearly are not the solutions of interest. 
	\item The mesh is always uniform and cannot be adjusted based on the physics. For example, in fluid dynamics, the mesh should be refined near the boundary to resolve the boundary layers.  
	\item Although a sufficiently high resolution can reduce representation errors and alleviate some of the aforementioned issues to a certain extent, it will significantly increase the training costs and cause memory issues. 
	\item Conventional rasterization/voxelization methods are not differentiable with respect to the input design parameters, limiting its use for geometry optimization and design purpose. 
\end{enumerate}

\subsection{Physics-Informed Geometry-Adaptive Convolutional Neural Network}
Motivated by the drawbacks and limitations of classic CNNs as discussed above, we propose a novel physics-informed geometry-adaptive CNN approach (PhyGeoNet), enabling CNN-based physics-informed learning for problems with non-uniform grids and irregular geometries. The key idea is to utilize coordinate transformation techniques to map solution fields from irregular physical domain to rectangular reference domain. Hence, the powerful classic uniform grid based CNN backbone can be directly leveraged by reformulating the physics-constrained optimization (Eq.~\ref{eqn:pdetraining}) on the reference domain. The geometry transformation is a deterministic process and can be precomputed, so it can be seen as a part of network architecture and no aliasing errors will be introduced. Moreover, various BCs are strictly enforced in a hard manner, which is in contrast to previous works, where the BCs are often treated as penalty terms and imposed softly. 

\subsubsection{Coordinate transformation between physical and reference domains}
The forward/inverse mapping between coordinates of the irregular physical domain ($\Omega_{p}$) and regular reference domain ($\Omega_{r}$) can be defined as, 
\begin{equation}
\label{eqn:GeoMap}
\mathbf{x}=\mathcal{G}(\boldsymbol{\xi}), \ \ 
\boldsymbol{\xi}=\mathcal{G}^{-1}(\mathbf{x}),
\end{equation}
where $\mathcal{G}:\Omega_{r}\mapsto\Omega_{p}$ denotes the forward map, and $\mathcal{G}^{-1}:\Omega_{p}\mapsto\Omega_{r}$ denotes the inverse map; $\mathbf{x}\in\Omega_{p}$ and $\boldsymbol{\xi}\in\Omega_{r}$ represent mesh coordinates of the physical domain and reference domain, respectively. Typically, non-uniform mesh grids are specified in the physical domain with arbitrary geometry, while the corresponding grids of the reference domain are uniformly spaced in a regular geometry (e.g., rectangular or cuboid). Given coordinate transformation functions ($\mathcal{G}$/$\mathcal{G}^{-1}$), the deterministic one-to-one mapping of coordinates from the reference domain to the physical domain can be uniquely determined and Jabobians of the transformation map are also available to reformulate the PDEs on the reference domain. However, in most cases, analytical forms of $\mathcal{G}$ or $\mathcal{G}^{-1}$ are not available, which have to be approximated numerically. Therefore, elliptic coordinate transformation~\cite{thompson1974automatic} is applied here, and the general idea is to solve a boundary-value problem (i.e., elliptic equations) since the boundary conditions are given in both physical and reference domains.

\begin{figure}[H]
	\centering
	{\includegraphics[width=0.8\textwidth]{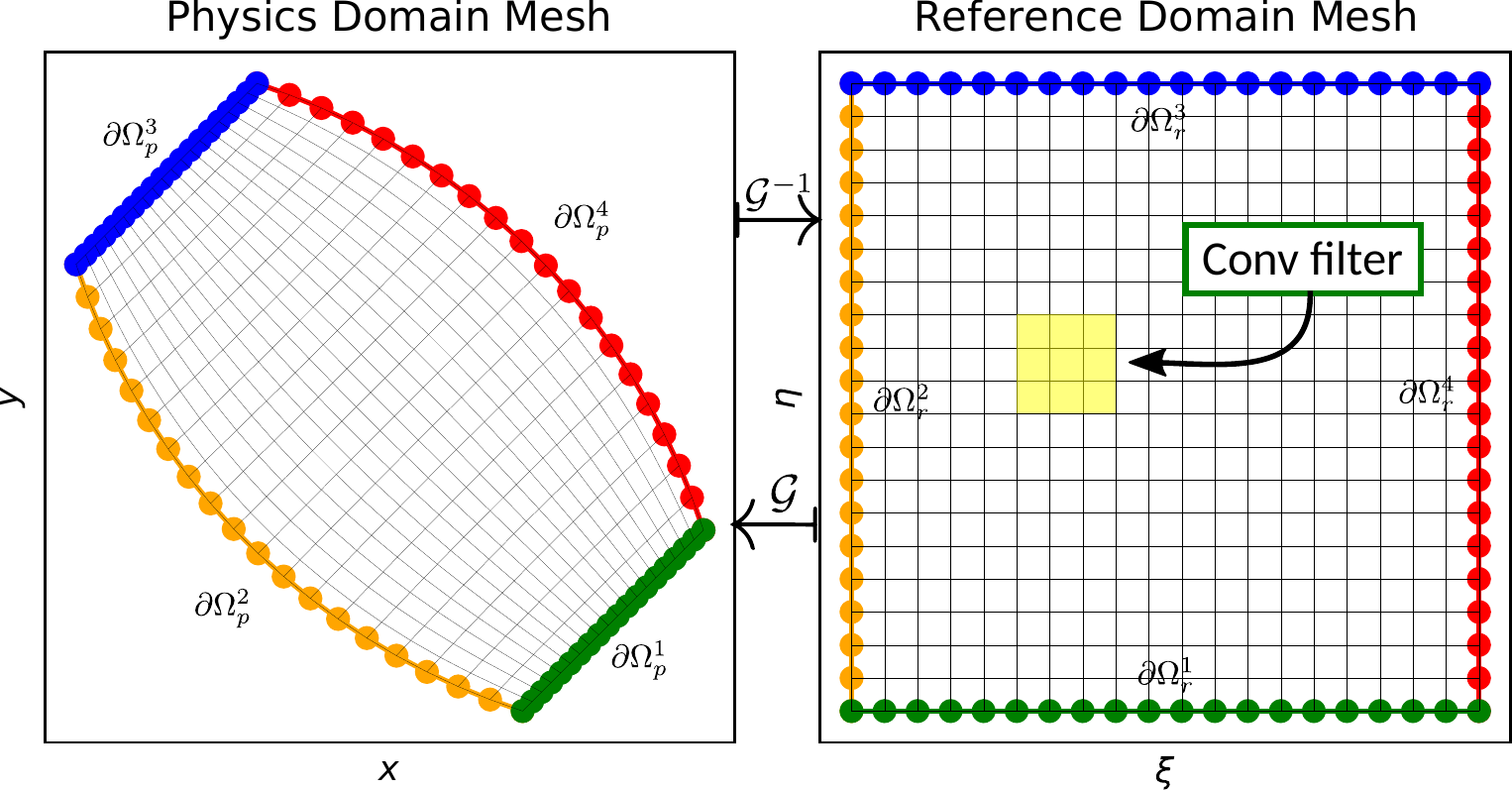}}
	\caption{A schematic diagram of the coordinate mappings between an irregular physical domain and regular reference domain. Classic convolution operations are only acted on the reference domain.}
	\label{fig:PhyGeoNetMap}
\end{figure}

Without loss of generality, two-dimensional (2-D) elliptic transformation is introduced, i.e., $\Omega_r, \Omega_p \subset \mathbb{R}^2$. Consider an irregular domain bounded by four edges $\partial\Omega_p^i, i = 1, \cdots, 4$ (Fig.~\ref{fig:PhyGeoNetMap}a), and the corresponding reference domain is a rectangular bounded by edges $\partial\Omega_r^i, i = 1, \cdots, 4$ (Fig.~\ref{fig:PhyGeoNetMap}b).
The coordinates of the physical domain and reference domains are denoted by $\mathbf{x}\doteq[x,y]$ and $\boldsymbol{\xi}\doteq[\xi,\eta]$, respectively. The one-to-one mapping on the boundary grids for physical and reference domain are usually known \emph{a priori}, i.e., 
\begin{linenomath*}
\begin{subequations}
	\label{eqn:GinvBC}
	\begin{alignat}{2}
	\boldsymbol{\xi}(\mathbf{x})&=\boldsymbol{\xi}_{b}\;\ \mathrm{for}\;\forall\;\mathbf{x}\in\partial\Omega_p^i, i = 1, \cdots, 4\\
	\mathbf{x}(\boldsymbol{\xi})&=\mathbf{x}_{b}\;\ \mathrm{for}\;\forall\;\boldsymbol{\xi}\in\partial\Omega_r^i, i = 1, \cdots, 4 
	\end{alignat}
\end{subequations}
\end{linenomath*}
where $\boldsymbol{\xi}_b$ and $\mathbf{x}_b$ are given. To numerically approximate $\mathcal{G}$ or $\mathcal{G}^{-1}$, one can solve an elliptic equations with specified boundary conditions in Eq.~\ref{eqn:GinvBC}. For example, the inverse map $\mathcal{G}^{-1}$ can be obtained by solving a diffusion equation,
\begin{equation}
\label{eqn:GInv}
\nabla^2\boldsymbol{\xi}(\mathbf{x}) = 0,
\end{equation}
with the BC defined in Eq.~\ref{eqn:GinvBC}a. However, the inverse map $\mathcal{G}^{-1}:\Omega_{p}\mapsto\Omega_{r}$ is less useful since the set of physical grids $\mathbf{x}$ that can be converted to a set of uniform grids are usually unknown. Instead, it is practical to solve the forward map $\mathcal{G}:\Omega_{r}\mapsto\Omega_{p}$, which locates corresponding collocation points in physical domain given uniform mesh grids in the reference domain. By interchanging the independent and dependent variables in Eq.~\ref{eqn:GInv}, the following diffusion equations in terms of physical coordinate $\mathbf{x}$ is derived,
\begin{linenomath*}
	\begin{subequations}
		\label{eqn:G}
		\begin{alignat}{2}
		\alpha\frac{\partial^2x}{\partial\xi^2}-2\beta\frac{\partial^2x}{\partial\xi\partial\eta}+\gamma\frac{\partial^2x}{\partial\eta^2}=0,\\
		\alpha\frac{\partial^2y}{\partial\xi^2}-2\beta\frac{\partial^2y}{\partial\xi\partial\eta}+\gamma\frac{\partial^2y}{\partial\eta^2}=0,
		\end{alignat}
	\end{subequations}
\end{linenomath*}
where $\alpha$, $\beta$, and $\gamma$ are given by,
\begin{equation}
\label{eqn:alphabetagamma}
\begin{split}
\alpha&=\left(\frac{\partial x}{\partial \eta}\right)^2+\left(\frac{\partial y}{\partial \eta}\right)^2,\\
\gamma&=\left(\frac{\partial x}{\partial \xi}\right)^2+\left(\frac{\partial y}{\partial \xi}\right)^2,\\
\beta&=\frac{\partial x}{\partial \xi}\frac{\partial x}{\partial \eta}+\frac{\partial y}{\partial \xi}\frac{\partial y}{\partial \eta}.
\end{split}
\end{equation}
By solving Eq.~\ref{eqn:G} with BC defined in Eq.~\ref{eqn:GinvBC}b numerically (i.e., iterative method), the discrete values of the forward map $\mathcal{G}$ are obtained. The proof that Eq.~\ref{eqn:G} holds given Eq.~\ref{eqn:GInv} are provided in \ref{sec:proofINVMap}.

\subsubsection{Reformulate physics-constrained learning on reference domain}
To enable physics-constrained learning in the reference domain, the original PDEs defined in the physical domain (Eq.~\ref{eqn:generalSteadyPDE}) has to be recast as the form in the reference domain. Specifically, Jacobians of the transformation map $\mathcal{G}$ are computed in order to convert differential operators from the physical domain to the reference domain. For instance, the first derivatives in $\Omega_{p}$ is transformed into $\Omega_{r}$ as,
\begin{linenomath*}
	\begin{subequations}
		\label{eqn:chainrule}
		\begin{alignat}{2}
		\frac{\partial}{\partial x}&=\left(\frac{\partial}{\partial\xi}\right)\left(\frac{\partial\xi}{\partial x}\right) + 
		\left(\frac{\partial}{\partial\eta}\right)\left(\frac{\partial\eta}{\partial x}\right),\\
		\frac{\partial}{\partial y}&=\left(\frac{\partial}{\partial\xi}\right)\left(\frac{\partial\xi}{\partial y}\right) + 
		\left(\frac{\partial}{\partial\eta}\right)\left(\frac{\partial\eta}{\partial y}\right).
		\end{alignat}
	\end{subequations}
\end{linenomath*}
Typically, finite differences are used to numerically calculate Jacobians, which has to be performed in the reference domain. Therefore, the inverse transformation is applied~\cite{anderson1995computational} to modify Eq.~\ref{eqn:chainrule} as,
\begin{linenomath*}
	\begin{subequations}
		\label{eqn:Du}
		\begin{alignat}{2}
	\frac{\partial }{\partial x}&=\underbrace{\frac{1}{J}}_{\text{constant}}\Big[\left(\frac{\partial }{\partial \xi}\right) \underbrace{\left(\frac{\partial y}{\partial \eta}\right)}_{\text{constant}} - \left(\frac{\partial }{\partial \eta}\right) \underbrace{\left(\frac{\partial y}{\partial \xi}\right)}_{\text{constant}}\Big],\\
	\frac{\partial }{\partial y}&=\underbrace{\frac{1}{J}}_{\text{constant}}\Big[\left(\frac{\partial }{\partial \eta}\right) \underbrace{\left(\frac{\partial x}{\partial\xi}\right)}_{\text{constant}} - \left(\frac{\partial }{\partial \xi}\right) \underbrace{\left(\frac{\partial x}{\partial \eta}\right)}_{\text{constant}}\Big],
		\end{alignat}
	\end{subequations}
\end{linenomath*}
where $J=\frac{\partial x}{\partial\xi}\frac{\partial y}{\partial\eta}-\frac{\partial x}{\partial\eta}\frac{\partial y}{\partial\xi}\neq0$ is the determinant of the Jacobian matrix and metrics $\frac{\partial y}{\partial\eta}$, $\frac{\partial y}{\partial\xi}$, $\frac{\partial x}{\partial\eta}$, and $\frac{\partial x}{\partial\xi}$ can be precomputed and remain constant if $\mathcal{G}$ is defined. 
Therefore, differential operators $\frac{\partial}{\partial\eta}$ and $\frac{\partial}{\partial\xi}$ are defined on the reference domain, which are implemented as classic CNN filters defined by finite difference (FD) stencils. To obtain the second or higher derivatives (e.g., $\frac{\partial^2}{\partial x^2}$ and $\frac{\partial^2}{\partial xy}$), the FD-based CNN filters (Eq.~\ref{eqn:Du}) are applied to the solution field for multiple times, instead of using chain rule (Eq.~\ref{eqn:chainrule}), which overcomplicates the implementation.  More details can be found in \ref{sec:filterDetail}. 

With these modified derivative terms, the optimization defined in Eq.~\ref{eqn:pdetraining} is reformulated on the reference domain as,
\begin{equation}
\label{eqn:optmizationNew}
\begin{split}
&\min_{\Gamma} \sum_{i=1}^{n_d}
\underbrace{\left\|
	\tilde{\mathcal{F}}\Big(\mathbf{u}^{cnn}(\boldsymbol{\Xi}, \boldsymbol{\mu}_i; \Gamma),
	\nabla\mathbf{u}^{cnn}(\boldsymbol{\Xi}, \boldsymbol{\mu}_i; \Gamma), 
	\nabla^2\mathbf{u}^{cnn}(\boldsymbol{\Xi}, \boldsymbol{\mu}_i; \Gamma), \cdots; \boldsymbol{\mu}_i\Big) \right\|_{\Omega_{r}}}_{\text{equation-based loss on reference domain:}\ \tilde{\mathcal{L}}_{pde}},\\
&s.t.\;\; \tilde{\mathcal{B}}\Big(\mathbf{u}^{cnn}(\boldsymbol{\Xi}, \boldsymbol{\mu}_i; \Gamma),
\nabla\mathbf{u}^{cnn}(\boldsymbol{\Xi}, \boldsymbol{\mu}_i; \Gamma), 
\nabla^2\mathbf{u}^{cnn}(\boldsymbol{\Xi}, \boldsymbol{\mu}_i; \Gamma), \cdots;
\boldsymbol{\mu}_i\Big)=0,\;\;\text{on }\partial\Omega_{r},
\end{split}
\end{equation}
where $\tilde{\mathcal{F}}$ and $\tilde{\mathcal{B}}$ are modified PDE and boundary operators on reference domain, and $\boldsymbol{\Xi}$ are discrete reference coordinates. 

\subsubsection{Hard enforcement of boundary conditions}
The solutions are uniquely determined by given BCs, and thus proper enforcement of BCs is essential when labeled data are absent, otherwise, the optimization problem becomes ill-posed. {In general, the constrained optimization defined in Eq.~\ref{eqn:optmizationNew} can be converted to an unconstrained one by treating BCs as penalty terms, known as the \emph{soft BC enforcement}. As an alternative, the contraints can be strictly enforced and BCs will be automatically satisfied by construction, kwnon as the \emph{hard BC enforcement}}  
\begin{figure}[H]
	\centering
	{\includegraphics[width=0.8\textwidth]{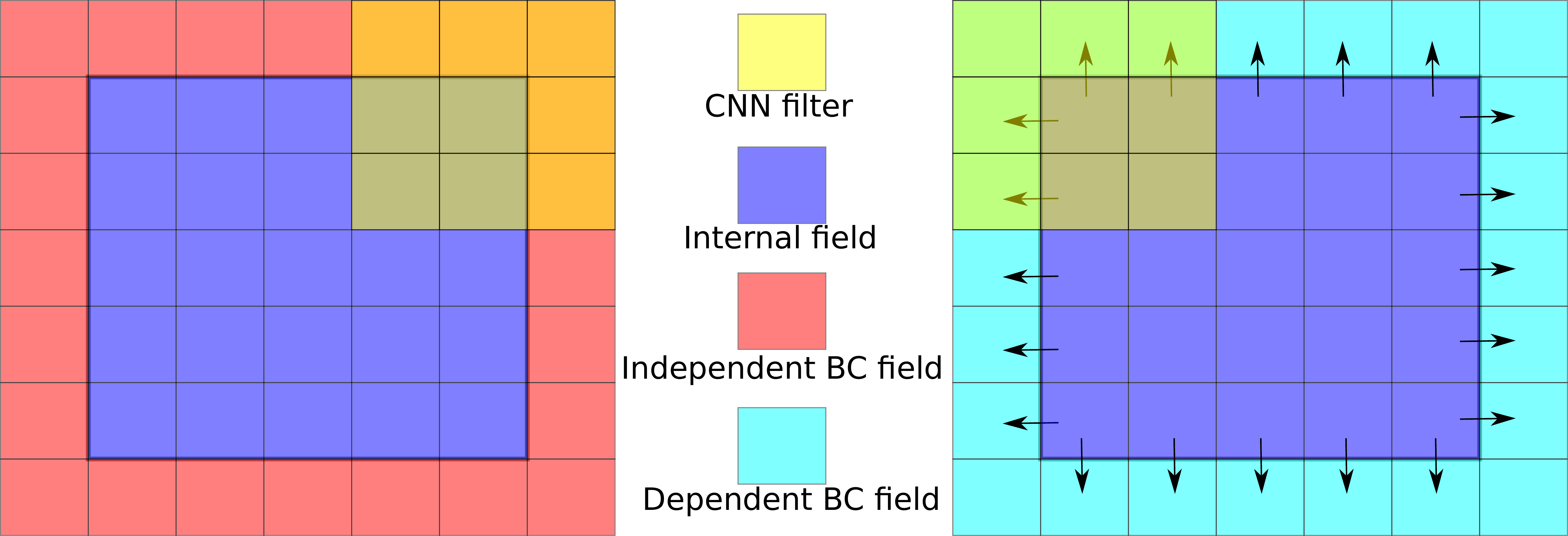}}
	\caption{Hard enforcement of BCs via padding. (left) Dirichlet BCs are strictly imposed by constant padding, which are fixed during training, and (right) Neumann BCs are imposed by padding that are derived from the internal field based on finite differences.}
	\label{fig:BCDemo}
\end{figure}
As discussed by Sun et al.~\cite{sun2020surrogate}, the soft BC enforcement has several drawbacks and slows the convergence. In this work, a hard BC enforcement is proposed, where both Dirichlet and Neumann BCs are strictly satisfied at the discrete level based on padding, referring to layers of pixels added to images when it is being processed by the CNN filters. As shown in Fig.~\ref{fig:BCDemo} (left), Dirichlet BCs can be exactly imposed by applying constant padding, which does not vary during training.  As for the Neumann BCs, the values of padding are derived from the solutions at internal nodes via finite differences, Although the padding values vary at each training epoch, the difference relations defined by Neumann BCs are strictly enforced.

\subsubsection{Physics-informed CNN architecture in reference domain}
The network architecture is built with input channels composed of physical parameters $\boldsymbol{\mu}$ and coordinates $\mathbf{x}$ of the physical domain. 
\begin{figure}[H]
	\centering
	{\includegraphics[width=1\textwidth]{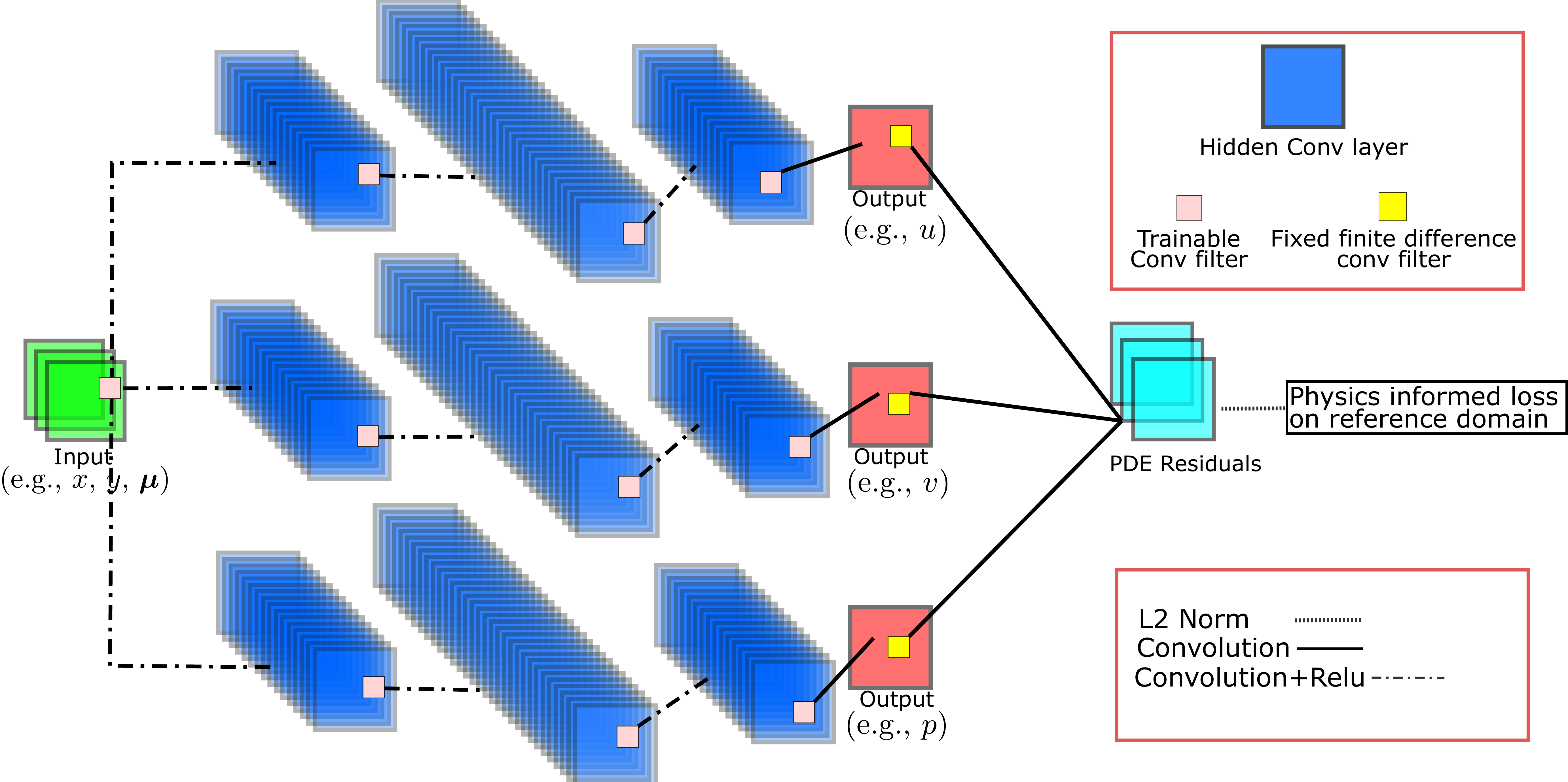}}
	\caption{The CNN architecture in reference domain, where forward propagation is from left to right. The solution field of each physical variable is described by a separate subnet.}
	\label{fig:CNNArch}
\end{figure}
The solution field of each state variable (e.g., pressure or velocity components) on the reference domain is represented by a separate sub-CNN, and thus the trainable network parameters (i.e., filters) are decoupled for different solution variables. This is because the values of different variables might be different in orders of magnitude (e.g., streamwise velocity component can be orders of magnitude greater than spanwise velocity component). The superiority of using separate subnets for multivariate regression has been demonstrated in both data-driven and data-free scenarios~\cite{han,sun2020surrogate}. Each sub-net has an identical structure of three hidden convolution layers, which is adapted from a classic CNN architecture proposed by Shi et al~\cite{shi2016real} originally for single image super-resolution. Specifically, each layer has a trainable filter with the same kernel size of $5 \times 5$, and 2-D conv operations with padding of 2 and stride of 1 are applied to ensure the output layer has the same size (height $\times$ width) as the input.  All the sub-nets are trained simultaneously with a unified physics-based loss function constructed by PDE residual layers using fixed finite difference filters. The trainable network parameters are initialized from a uniform distribution of $\mathcal{U}\left(-\sqrt{\frac{1}{25C_{in}}},\sqrt{\frac{1}{25C_{in}}}\right)$, where $C_{in}$ is number of input channels. Fig.~\ref{fig:CNNArch} gives an overview of the proposed network architecture.

\section{Numerical Results\label{sec:result}}
In this section, several elliptic and parabolic PDE systems are studied to demonstrate the merit of the proposed PhyGeoNet for label-free learning on irregular geometries. To begin with, we first present PhyGeoNet solutions of the {non-parametric} heat equation and Navier-Stokes equations on irregular domains with arbitrary shapes. Then, the experiments on more challenging cases with varying parameters are explored. Specifically, {the solutions of the heat equation with parameterized Dirichlet boundary conditions, Navier-Stokes equations with varying geometries, and Poisson equations with spatially-varying source term are learned without any data}. To evaluate the learning/prediction performance, finite volume (FV)-based numerical simulations are conducted as the benchmark for comparisons, and the relative error metric is defined as,
\begin{equation}
\mathrm{Error}=\sqrt{\frac{||\tilde{\mathbf{u}}-\mathbf{u}^{cnn}||_{L^2}}{||\tilde{\mathbf{u}}||_{L^2}}},
\label{eqn:errormetric}
\end{equation}
where $\tilde{\mathbf{u}}$ is the ground truth (FV solution), $\mathbf{u}$ is the PhyGeoNet predictions, and $||\cdot||_{L^2}$ represents the L2 norm. The FV simulations are performed on an open-source FV platform, OpenFOAM \cite{jasak2007openfoam}. {In particular, the convection term is discretized based on Gauss theorem with second-order bounded linear upwind interpolation (i.e., Gauss linearUpwind Scheme in OpenFOAM), and the diffusion term is discretized using central Gauss linear interpolation with the explicit non-orthogonal correction scheme for the surface normal gradient (i.e., Gauss linear corrected). The Semi-Implicit Method for Pressure Linked Equations (SIMPLE) algorithms were used for solving the incompressible Navier-Stokes equations~\cite{pletcher2012computational}, and the Rhie and Chow interpolation with collocated grids was employed to prevent the pressure–velocity decoupling~\cite{rhie1983numerical}.} The proposed PhyGeoNet is implemented in PyTorch~\cite{paszke2017automatic} and the training is conducted on an NVIDIA GeForce RTX 2080 Graphics Processing Unit (GPU) card. The computational parameter settings for all cases are summarized in Tab.~\ref{tab:ParamterSummary} (See \ref{Appendix:HPS}). The code and datasets for this work will become available at \url{https://github.com/Jianxun-Wang/phygeonet} upon publication.

\subsection{{Non-parametric solutions}}
\subsubsection{Heat equation (Case 1)}
\label{sec:detHeat}
\begin{figure}[htp]
	\centering
	{\includegraphics[width=0.8\textwidth]{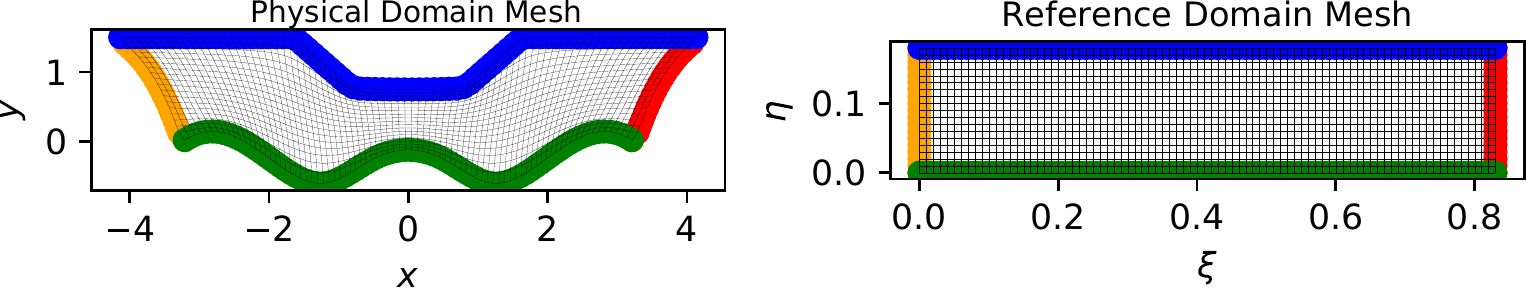}}
	\caption{The physical domain (left) and reference domain (right) for {non-parametric} heat equation. The corresponding edges of two domains are indicated by the same color.}
	\label{fig:DetHeatDomain}
\end{figure}

\begin{figure}[htp]
	\centering
	{\includegraphics[width=0.8\textwidth]{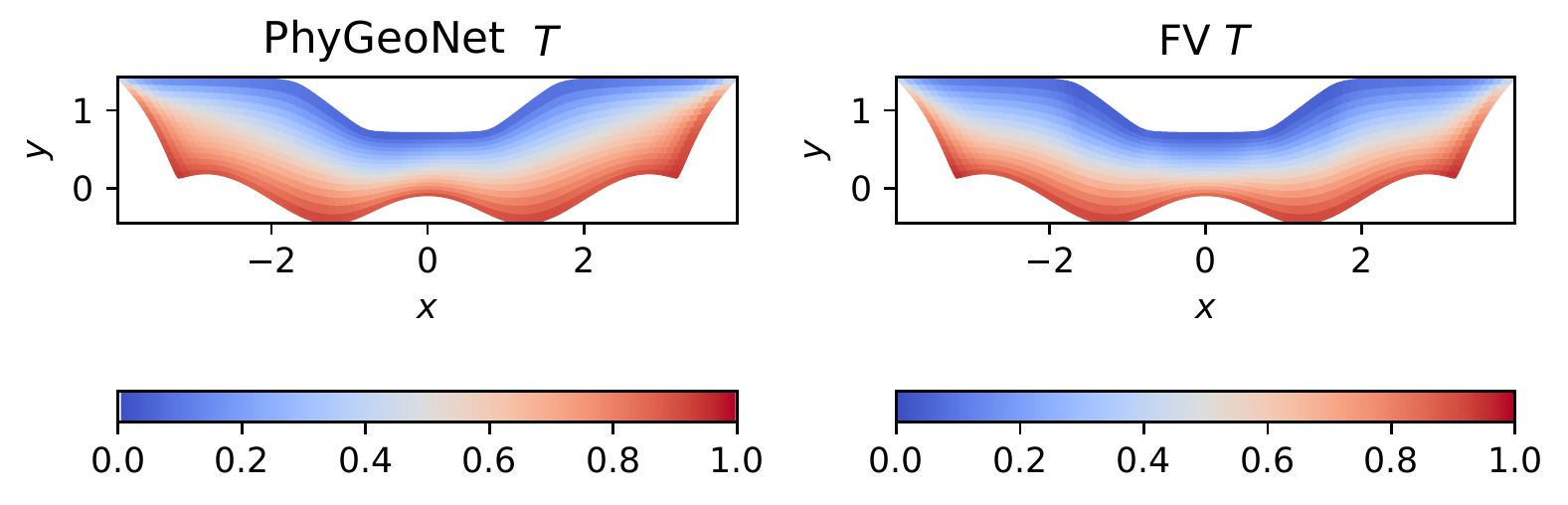}}
	\caption{Comparison of temperature fields obtained by PhyGeoNet and FV-based simulations.}
	\label{fig:DetHeatSol}
\end{figure}

Prediction of temperature distribution across an irregular surface requires to solve the heat equation $\Omega_p$, which is of importance to industrial design of, e.g., chip cooling and panel heating~\cite{miriel2002radiant,chowdhury2009chip}. Here we consider the heat equation defined as,
\begin{equation}
\begin{split}
\nabla\cdot (\nabla T(\mathbf{x})) =0, \  \mathbf{x}\in\Omega_p,\\
T(\mathbf{x}) = T_{bc}(\mathbf{x}), \ \mathbf{x}\in\partial\Omega_p,
\end{split}
\label{eqn:heateqn}
\end{equation}
where $T$ is temperature field and Dirichlet BC $T_{bc}$ is given. The physical domain $\Omega_{p}$ has an irregular shape with boundary-fitted mesh as shown in Fig.~\ref{fig:DetHeatDomain} (left panel), for which PDE-constrained learning is unable to be formulated using classic CNNs~\cite{long2019pde,zhu2019physics,mo2020integration,geneva_modeling_2020}. In our proposed PhyGeoNet, the PDE-based loss function and network optimization are defined in the regular reference domain (right panel of Fig.~\ref{fig:DetHeatDomain}), which is mapped from the irregular physical domain using elliptic transformation, and thus the solution field can be learned without any labeled data.
Fig.~\ref{fig:DetHeatSol} shows the comparison between the PhyGeoNet-predicted $T$ field and FV-based solution (benchmark). 
It can be seen that the temperature on the upper boundary (blue) is set as $T=0$, while on all other sides (green, red, and orange) $T=1$. Compared to the FV benchmark, the heat distribution can be accurately captured by the PhyGeoNet with the relative prediction error of $0.098$, demonstrating that the {non-parametric} heat equation can be solved by PhyGeoNet on irregular domains. The learning curve for this case is plotted in Fig.~\ref{fig:trainingloss}a. 

\subsubsection{Navier-Stokes equations (Case 2)}
\label{sec:N-S}
\begin{figure}[htp]
	\centering
	{\includegraphics[height=0.40\textwidth]{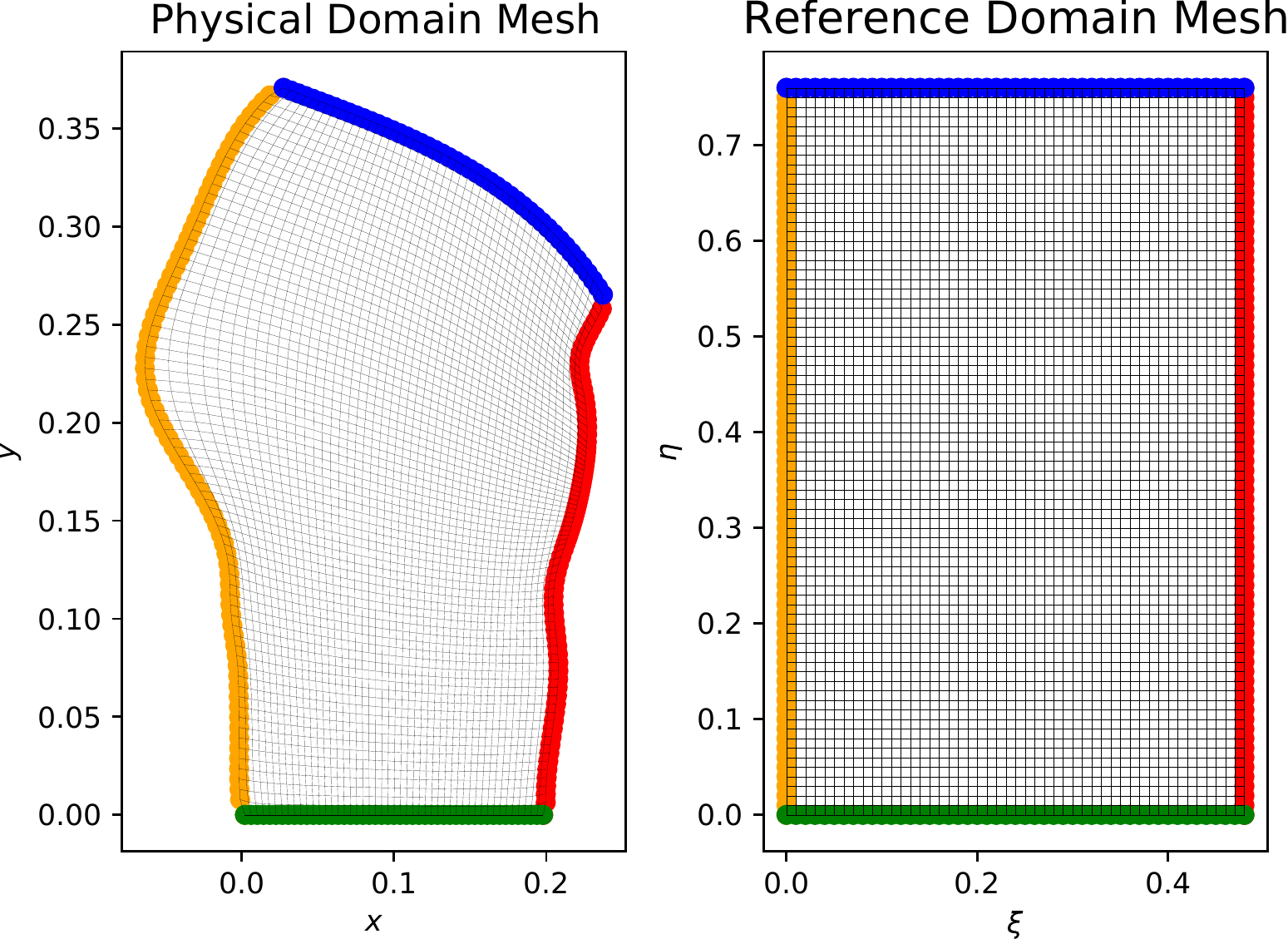}}
	\caption{The physical domain (left) and reference domain (right) for {non-parametric} Navier-Stokes equations. The corresponding boundary edges of two domains are indicated by the same color.}
	\label{fig:DetNSDomain}
\end{figure}

There has been growing interests in using CNNs to simulate/model fluid flows governed by Navier-Stokes equations, but most of them are formulated in a data-intensive supervised fashion~\cite{kim2019deep,bhatnagar2019prediction,xu2019multi}. Even so, the representation of irregular geometries is challenging and has often to rely on pixel/voxel-based approaches using signed distance functions or descriptor distance functions, which may introduce considerable aliasing errors. Here, our proposed PhyGeoNet is trained in an unsupervised manner to learn the solutions of the steady incompressible Navier-Stokes equations,
\begin{linenomath*}
	\begin{subequations}
		\label{eqn:ns}
		\begin{alignat}{2}
		\nabla \cdot \mathbf{v} &=0,\\
		(\mathbf{v}\cdot\nabla)\mathbf{v} &=\nabla\cdot(\nu\nabla\mathbf{v})-\nabla p,
		\end{alignat}
	\end{subequations}
\end{linenomath*}
where $\mathbf{v}$ is velocity vector, $p$ is pressure, and $\nu$ is fluid viscosity. The fluid flow is solved on an irregular domain with a 2-D vascular geometry, as shown in Fig.~\ref{fig:DetNSDomain}. 
The boundary condition on the lower edge (green) is set as the inlet with a constant inflow $\mathbf{v}=[0,1]$. The no-slip wall boundary condition is prescribed on the left and right boundaries (orange and red). The outlet on the upper boundary (blue) is given by $\nabla \mathbf{v}\cdot\mathbf{n}=0$ and $p=0$, where $\mathbf{n}$ is the local wall-normal vector. The input to the PhyGeoNet only includes geometry (i.e., coordinates) in the physical domain.

\begin{figure}[htp]
	\centering
	\subfloat[{$Re=20$}]
	{\includegraphics[width=1\textwidth]{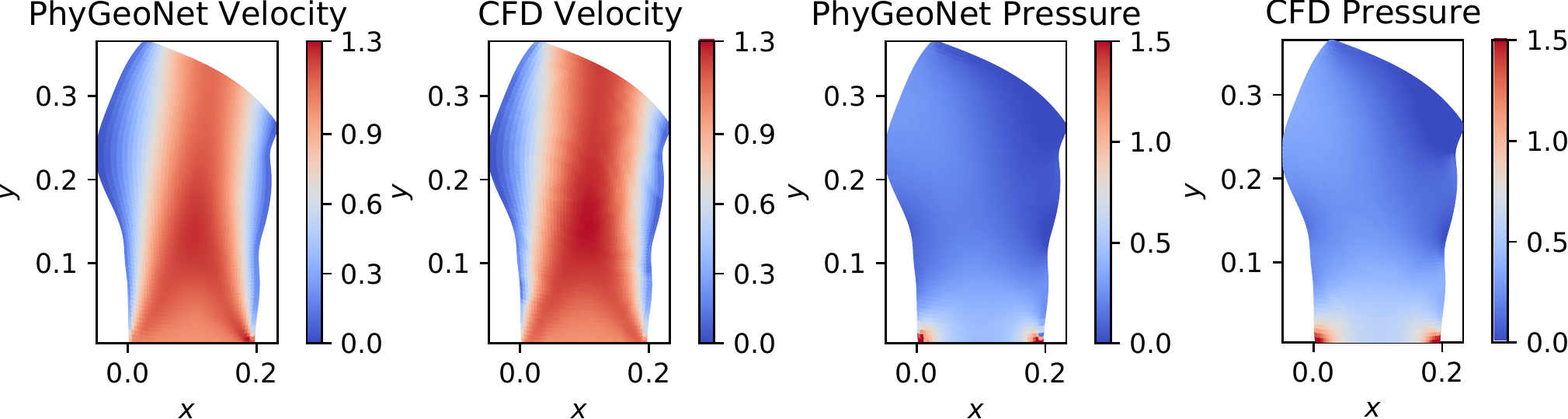}}
	\vfill
	\subfloat[{$Re=100$}]
	{\includegraphics[width=1\textwidth]{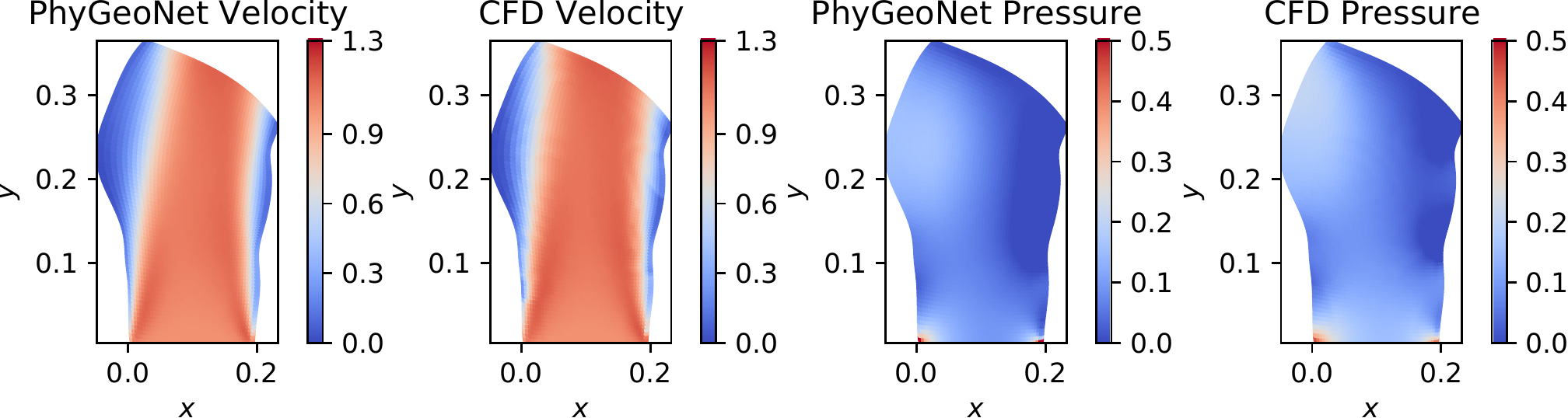}}
	\vfill
	\subfloat[{$Re=250$}]
	{\includegraphics[width=1\textwidth]{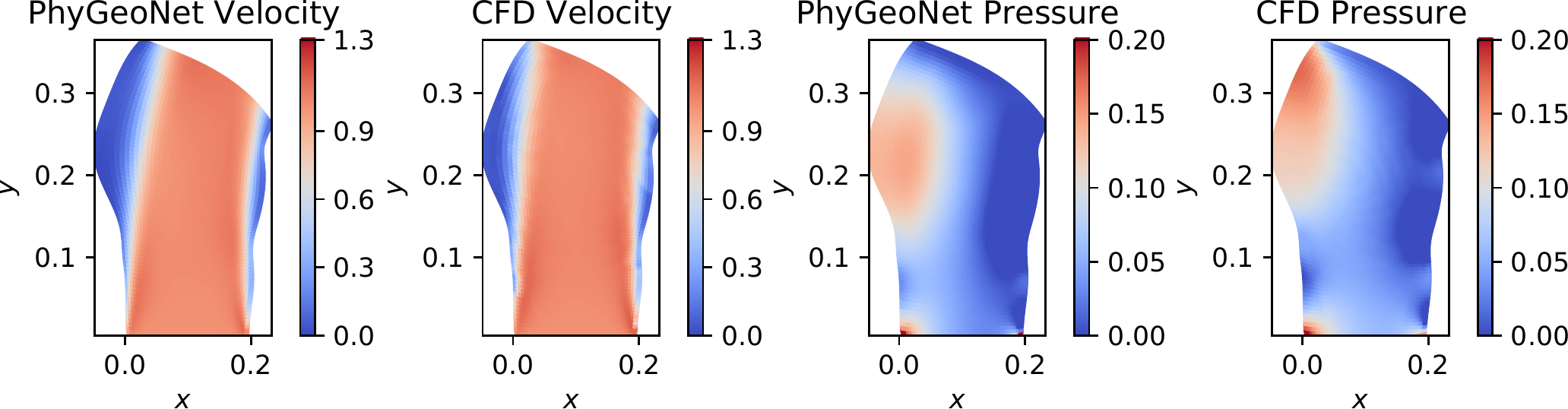}}
	\caption{The comparison between PheGeoNet and CFD solutions of velocity (left column) and pressure (right column).}
	\label{fig:DetNSSol}
\end{figure}

 The PhyGeoNet predicted solutions are shown in Fig.~\ref{fig:DetNSSol}, where the corresponding FV-based CFD results are plotted for comparison. 
 From the inlet, the fluid flow starts to move up towards the outlet. The boundary layers are gradually developed on both sides and velocity magnitude decreases close to the wall due to the friction between fluid and walls. Because of the conservation of mass, the magnitude of the center velocity is greater than that of the inlet velocity. As for the pressure, both left and right corners are relatively high-pressure regions, and the pressure is dropping down as the flow approaches the outlet due to the conservation of momentum. The curvature of the left wall located at the upper left of the domain bends the flow to the right slightly, which also leads to a relatively high-pressure region. Overall, PhyGeoNet is able to capture the flow field reasonably well compared to the CFD benchmark {for different Reynolds numbers in laiminar regime}. The relative errors for velocity magnitude and pressure are listed in Tab.~\ref{tab:DetNSErrorTab}, which provides a quantitative assessment of the performance. {We can see that the solution error slightly grows as the Reynolds number (Re) increases, which is because the convection becomes more dominant, posing challenges on PDE-driven learning based on central difference filters. Special numerical treatment, e.g., upwind differencing, may improve the results in this regard.} The learning curve (history of equation residuals) at Re = 20 is given in Fig.~\ref{fig:trainingloss}b. 
\begin{table}[H]
	\centering
	\begin{tabular}{||c c c||} 
		\hline
		Variable & velocity  magnitude& pressure \\ 
		\hline
		Relative  error {($Re=20$)} &  0.0783 & 0.347  \\ 
		Relative  error {($Re=100$)} &  0.0858 & 0.340  \\
		Relative  error {($Re=250$)} &  0.1150 & 0.410  \\
		\hline
	\end{tabular}
\caption{Relative error of learned solution for {non-parametric} Navier-Stokes equations.\label{tab:DetNSErrorTab}}
\end{table}

\subsection{Parametric solutions}
The real power of DNN-based approaches lies in learning PDE solutions in a parametric setting, since the trained DNN can be used as a surrogate model for rapid many-query applications. {To be specific, the PhyGeoNet will take the PDE parameters $\boldsymbol{\mu}$ as a part of the input and output the corresponding solution $\mathbf{u}(\boldsymbol{\mu})$, approximating the parameter-to-solution mapping $f: \boldsymbol{\mu}\mapsto \mathbf{u}$. To learn this mapping, the network will be trained on randomly selected parameter points $\{ \boldsymbol{\mu} \}_{train}$ by minimizing the equation residuals as defined in Eq.~\ref{eqn:optmizationNew}, where no labels are required. Therefore, the training set $\{ \boldsymbol{\mu} \}_{train}$ only includes input parameter points instead of output labels, which is in contrast to traditional data-driven DL models. After being fully trained, the PhyGeoNet can take new parameters $\{ \boldsymbol{\mu} \}_{test}$ to rapidly predict the corresponding solutions nearly without additional cost. In this subsection, we will demonstrate this capability of the PhyGeoNet for learning and predicting solutions of PDEs with varying parameters/boundary conditions.} 

\subsubsection{Heat equation with parameterized boundary conditions (Case 3)} 
\begin{figure}[htp]
	\centering
	{\includegraphics[width=0.6\textwidth]{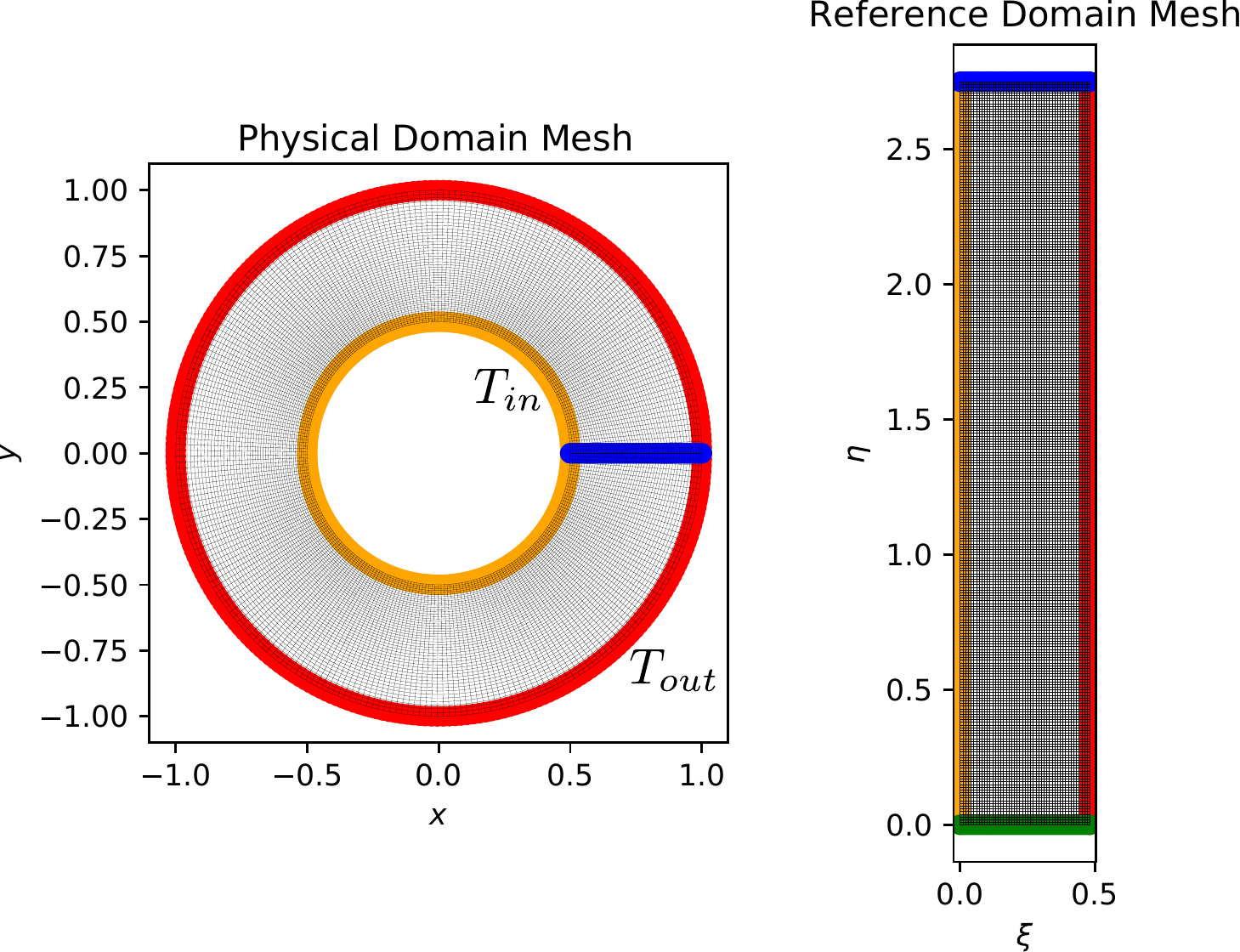}}
	\caption{The physical domain (left) and reference domain (right) for parametric heat equation. The corresponding boundary edges of two domains are indicated by the same color. Periodic boundary conditions are imposed on green and blue edges.}
	\label{fig:ParaHeatDomain}
\end{figure}

In this subsection, the PhyGeoNet is trained to learn the parametric solutions of the heat equation, where the boundary condition is varying.  A more complex irregular domain of a different topology is investigated here. Unlike the two geometries studied above, which both belong to the simply-connected domain, the annulus shape considered here (Fig.~\ref{fig:ParaHeatDomain}) is a doubly-connected domain \cite{spanier1989algebraic,rudin2006real}, which can be seen a Riemann surface,
\begin{equation}
 r<\big|\mathbf{x}-\mathbf{c}\big|<R,
\label{eqn:annlus}
\end{equation}
where $\mathbf{c}$ is the center of the annulus and $r$ and $R$ are inner and outer radii, respectively. To solve this problem, the doubly-connected domain is transformed back to a simply-connected domain by cutting off the annulus and imposing periodic boundary conditions at where it is cut (see Fig.~\ref{fig:ParaHeatDomain}).
The boundary temperature on the inner circle is parameterized as a varying constant of $T_{in}\in[1,7]$, while for outer circle boundary, a fixed temperature of $T_{out}=0$ is imposed. Now, the PhyGeoNet is built as a surrogate, whose input should reflect variations of boundary conditions. {In contrast to the non-parametric cases, where only the domain coordinates are provided to the network as the input, the PhyGeoNet for parametric cases should include both the geometry information (i.e., coordinates) and variable parameters. In this case, the field of a linear interpolation from the inner boundary $T_{in}$ to the outer one $T_{out}$ is provided as the input to the PhyGeoNet, which uniquely represents the variable boundary conditions.}

\begin{figure}[H]
	\centering
	{\includegraphics[width=0.8\textwidth,height=0.28\textwidth]{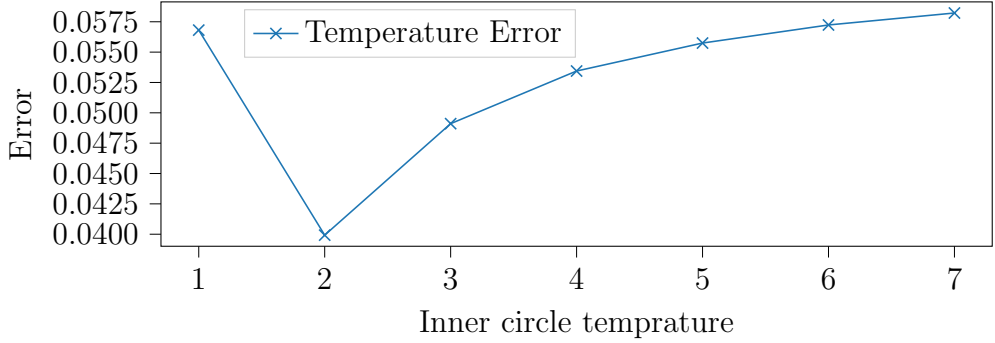}}
	\caption{Relative errors of PhyGeoNet predictions of different inner BCs.}
	\label{fig:ParaHeatError}
\end{figure}

{Unlike data-driven learning algorithms that are constrained by data availability}, the PhyGeoNet can be trained at any input points (i.e., $T_{in}$ here) due to its label-free nature.
\begin{figure}[htp]
	\centering
	{\includegraphics[width=0.55\textwidth]{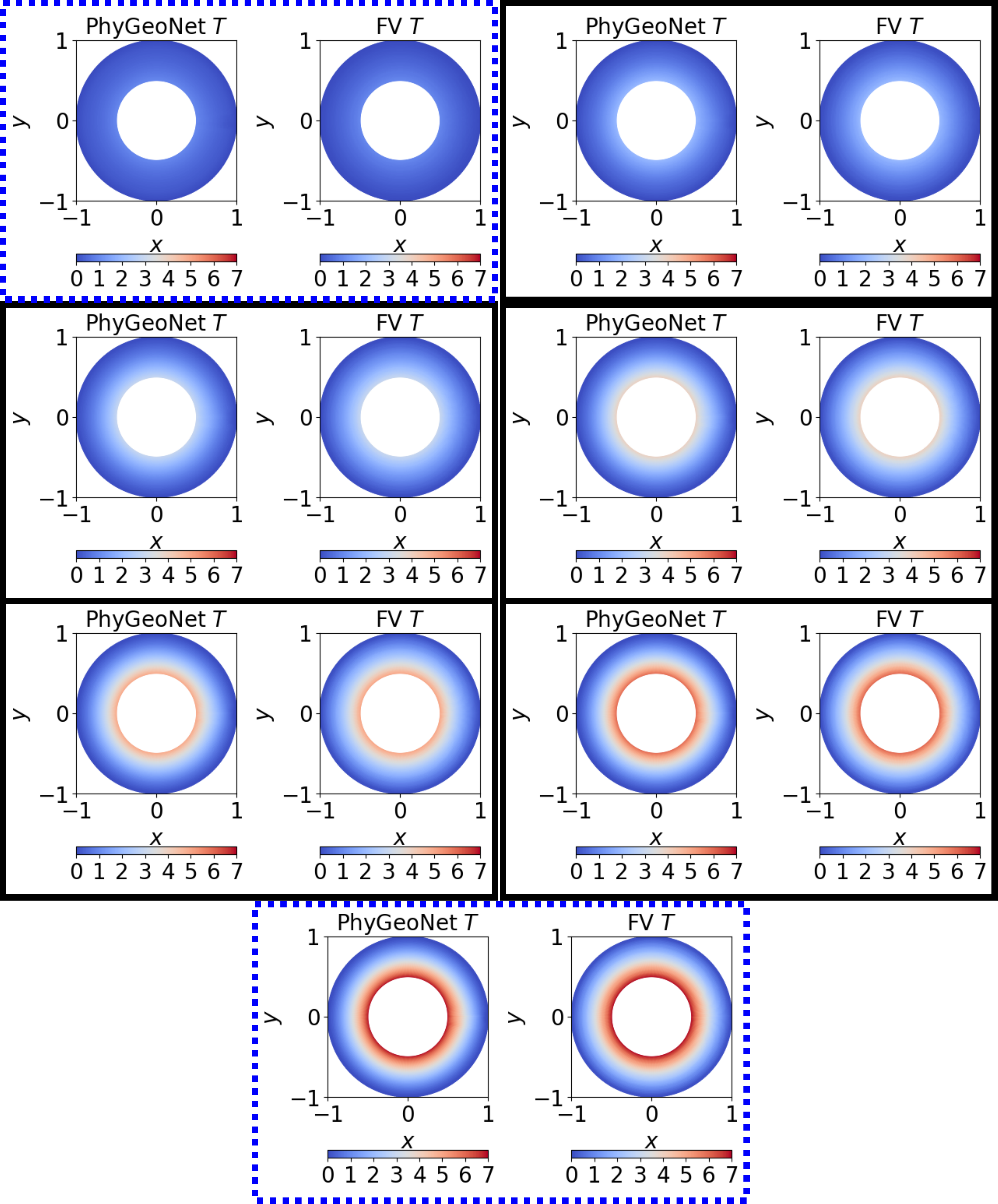}}
	\caption{The inner circle boundary temperature ($T_{in}$) is varying in $[1, 7]$ with a interval of $\delta T_{in}=1$. For cases in blue dashed blocks, the PhyGeoNet is trained in a PDE-driven fashion without data, while cases in black solid blocks are pure test points without training.}
	\label{fig:ParaHeatResult}
\end{figure}
Even so, it is still interesting to investigate the prediction performance at unseen input points, where PDE residuals have not been minimized. Therefore, the PhyGeoNet is trained only on two input samples ($T_{in}=1$ and $7$) in an unsupervised fashion and evaluated at seven parameter points across $T_{in}=[1, 7]$. The learning curve, i.e., convergence history of the PDE-based loss, is plotted in Fig.~\ref{fig:trainingloss}c. Fig.~\ref{fig:ParaHeatResult} shows the PhyGeoNet predictions compared with the FV-simulated benchmarks. The high temperature from the inner circle boundary is gradually diffused to zero towards the outer circle boundary. For all training and (unseen) test inner temperatures ($T_{in}$), the PhyGeoNet-predicted temperature contours agree with the numerical benchmarks from FV-based simulations very well. The relative errors of predictions at different parameters are lower than 0.05, as shown in Fig.~\ref{fig:ParaHeatError}. This demonstrates the potential of using PhyGeoNet as a surrogate for fast forward propagation of unseen input scenarios by leveraging strong expressibility and universal approximation capability of neural networks.

\subsubsection{Navier-Stokes equations with varying geometries (Case 4)\label{sec:parans}}
\begin{figure}[H]
	\centering
	{\includegraphics[width=0.8\textwidth]{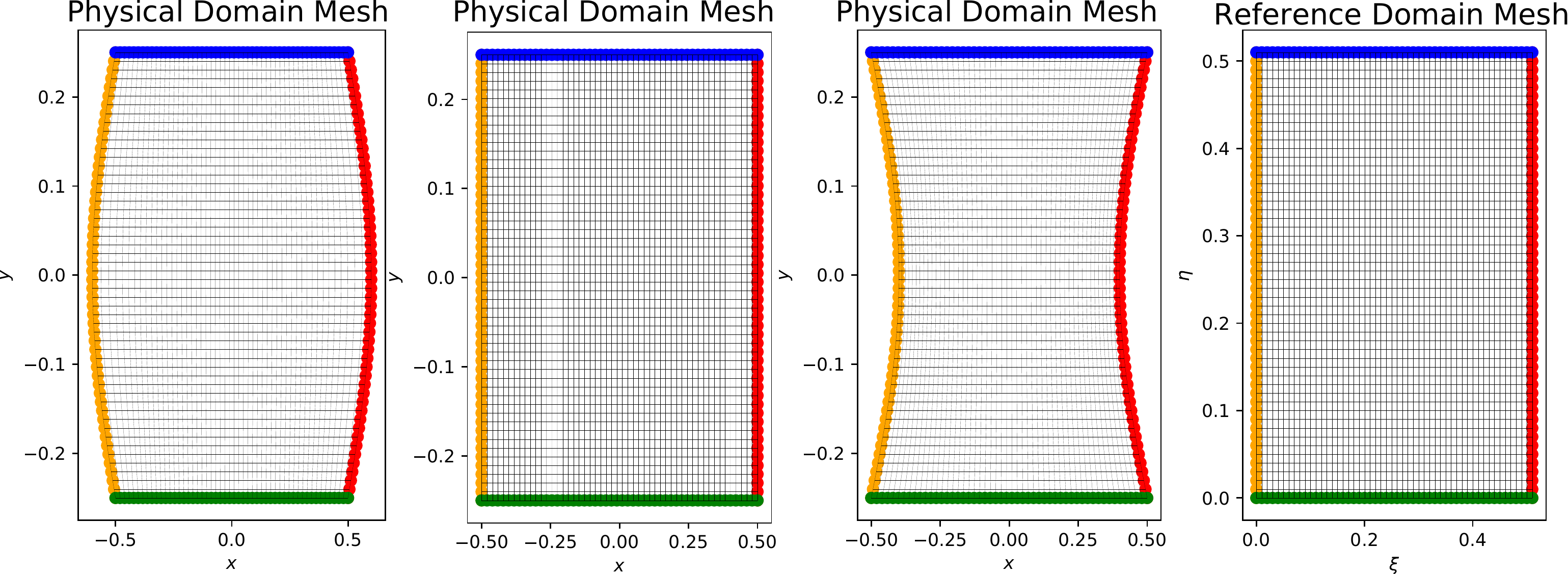}}
	\caption{From left to right: the first is aneurysm ($s=-0.1$), the second is neutral ($s=0.0$), the third is stenosis ($s=0.1$), and the last is the reference domain.}
	\label{fig:ParaNSTrainMesh}
\end{figure}

In the {second} numerical experiment, the proposed PhyGeoNet is applied for surrogate modeling of laminar flows with varying geometries, which is significant in many different applications. For example, in the context of image-based cardiovascular flow modeling, medical images have started to be widely utilized to construct geometry models for downstream CFD simulations, which, however, usually introduces large uncertainties due to image noise, segmentation errors, and other artifacts. To quantify and propagate geometric uncertainty in simulated quantities of interests (QoIs) of the blood flow solutions are critical but often require a massive number of forward CFD simulations~\cite{bozzi2017uncertainty}. Although various surrogate models are proposed to reduce the computational burden~\cite{gao2019bi,fleeter2019multilevel}, a considerable amount of labeled training data are still required. Here, the PhyGeoNet is employed to tackle this challenge without using any labels (i.e., CFD simulation data). Specifically, the steady incompressible Navier-Stokes equations (Eq.~\ref{eqn:ns}) with parameterized vascular geometries are solved, and the trained CNN is able to rapidly predict flow velocity and pressure fields for new geometries. Here idealized geometries of vascular stenosis and aneurysm are parameterized as,
\begin{equation}
\label{eqn:NSGeoPara}
\begin{split}
x_l=&s\cdot\cos{(2\pi y_l)}-0.5,\\
x_r=&-s\cdot\cos{(2\pi y_r)}+0.5,
\end{split}
\end{equation}
where $(x_l, y_l)$ and $(x_r, y_r)$ are coordinates of left and right vessel walls with $y_l, y_r \in[-0.25,0.25]$; $s\in[-0.1,0.1]$ is a scalar controlling the degree of stenosis or aneurysm. The top and bottom boundaries are defined as $(x_t,y_t)\in[-0.5,0.5]\times\{0.25\}$ and $(x_b,y_b)\in[-0.5,0.5]\times\{-0.25\}$. As shown in Fig.~\ref{fig:ParaNSTrainMesh}, the geometry of the physical domain can be controlled by changing the scalar parameter $s$ ($s>0$ for stenosis and $s<0$ for aneurysm). The boundary conditions of pressure and velocity are set as follows: the velocity inlet of $\mathbf{u}=[0,0.4]$ and $\nabla p \cdot \mathbf{n}=0$ is imposed at the lower edge (green); no-slip wall BC ($\mathbf{u}=[0,0]$ and $\nabla p\cdot\mathbf{n}=0$) is prescribed on the curved boundaries of both sides; the outlet is defined at the upper edge with $\nabla \mathbf{u}\cdot\mathbf{n}=0$ and $p=0$. Again, $\mathbf{n}$ is the local wall-normal vector.

\begin{figure}[H]
	\centering
	{\includegraphics[width=0.8\textwidth]{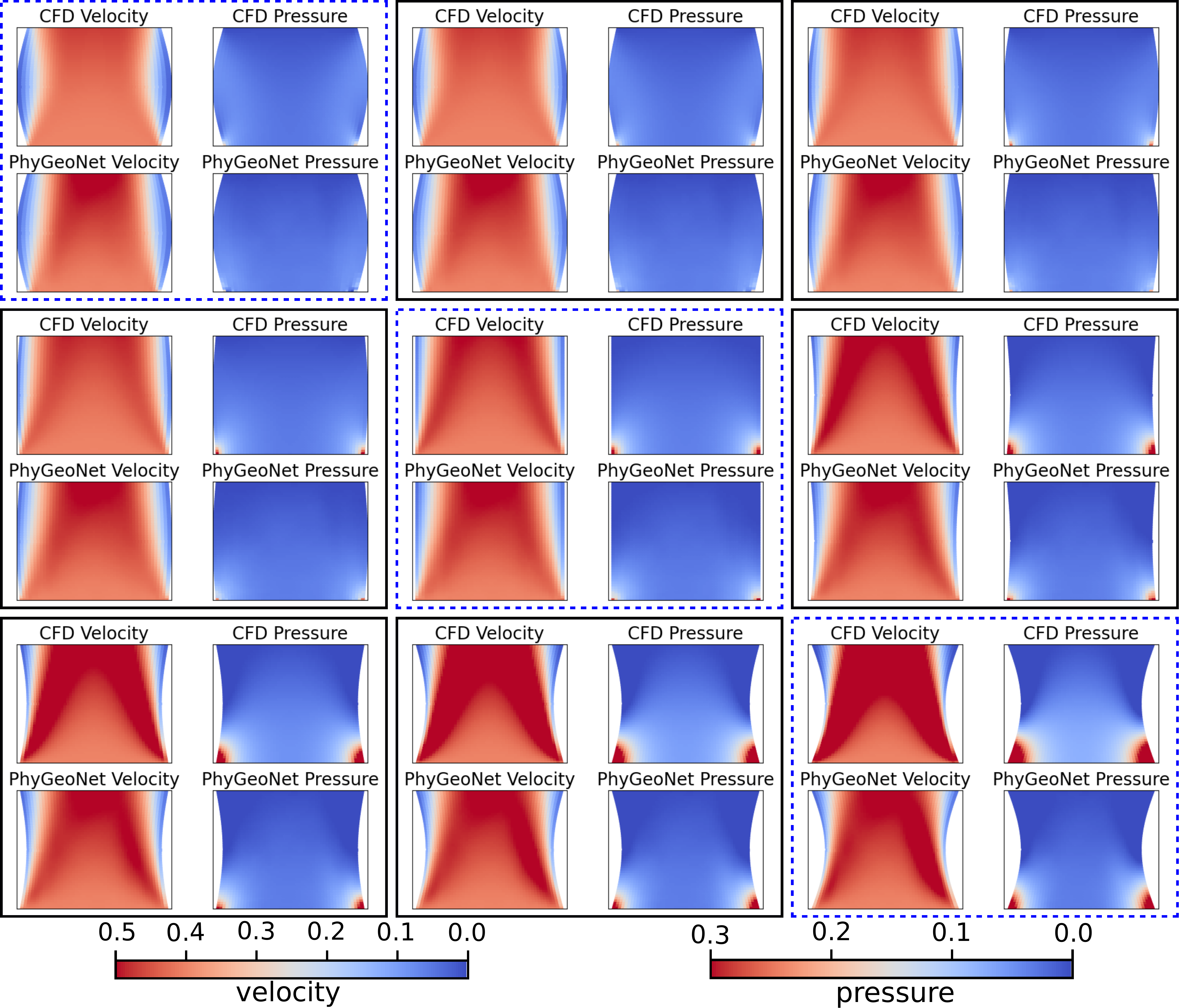}}
	\caption{The velocity and pressure contours from aneurysm ($s=-0.1$) to stenosis ($s=0.1$) with interval of $\delta s=0.025$ ({$Re=20$}). For cases in blue dashed blocks, the PhyGeoNet is trained in a PDE-driven fashion without data, while cases in black solid blocks are pure test points without training.}
	\label{fig:DetNSContour}
\end{figure}

\begin{figure}[H]
	\centering
	{\includegraphics[width=0.8\textwidth]{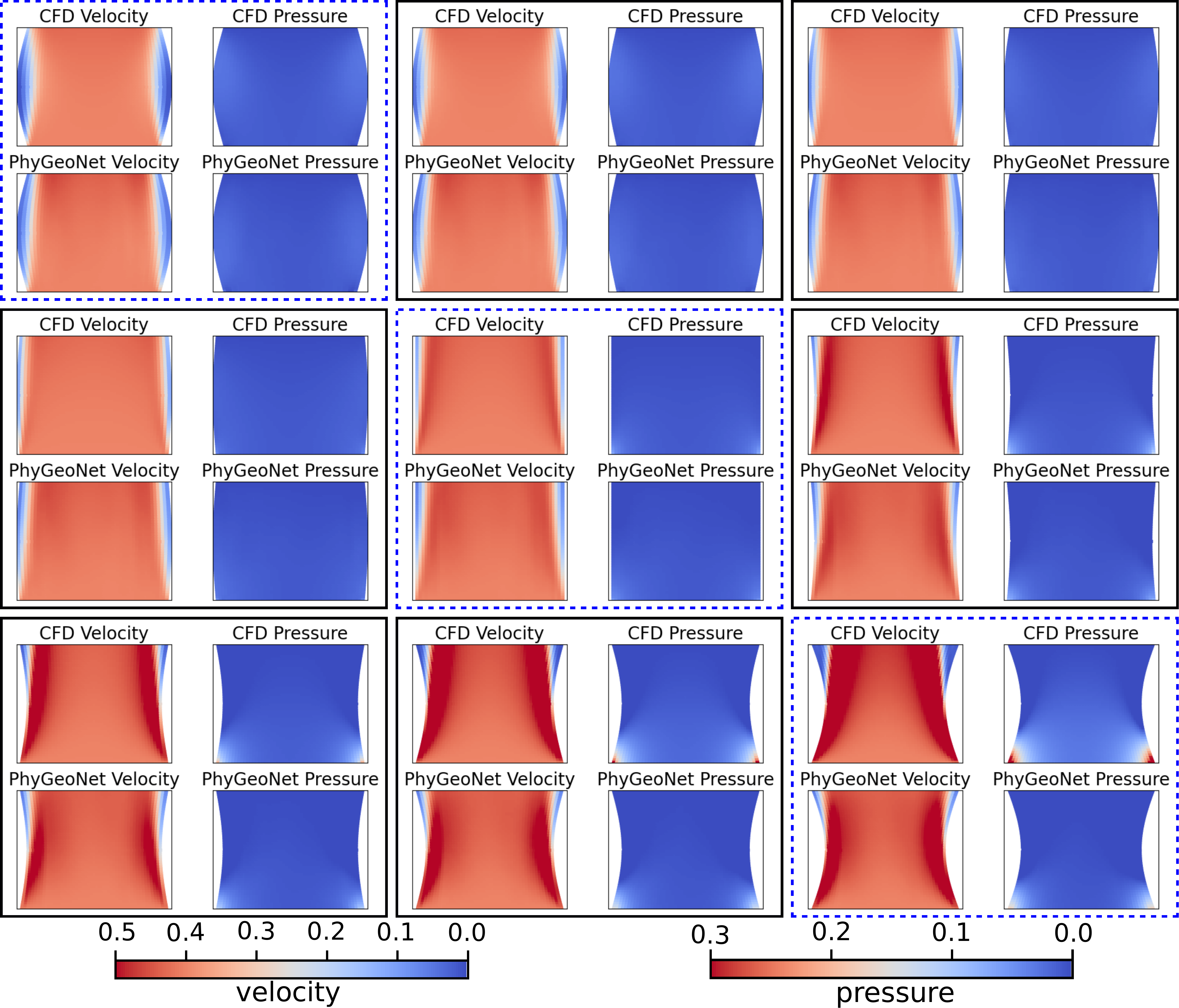}}
	\caption{{The velocity and pressure contours from aneurysm ($s=-0.1$) to stenosis ($s=0.1$) with interval of $\delta s=0.025$ ($Re=200$). For cases in blue dashed blocks, the PhyGeoNet is trained in a PDE-driven fashion without data, while cases in black solid blocks are pure test points without training.}}
	\label{fig:DetNSContour_Re200}
\end{figure}

Given that the physical domain is varying in shape, coordinates of the physical domain is used as the input for the PhyGeoNet to indicate different vascular geometries. Similarly, the Navier-Stokes-driven (Eq.~\ref{eqn:ns}) training is only conducted on three geometries as shown in blue dashed blocks in {Figs.~\ref{fig:DetNSContour} and~\ref{fig:DetNSContour_Re200}, and the trained modeled are tested by six different new {interpolated} geometries. The velocity and pressure contour plots for Re = 20 and 200 are presented in Figs.~\ref{fig:DetNSContour} and~\ref{fig:DetNSContour_Re200}, respectively.} The learning curves and physics-based loss convergence histories are given in Fig.~\ref{fig:trainingloss}d. 
\vspace{-1em}For all cases, the PhyGeoNet predictions generally agree with the CFD benchmarks and the contours are visually close to each other. Although no training data is used, the PhyGeoNet can capture the boundary layer variations due to the geometry changes, high-pressure regions located at the two lower corners, and the velocity acceleration near the outlet for all different shapes reasonably well.

\begin{figure}[ht]
	\centering
	\subfloat[{$R_e=20$}]
	{\includegraphics[width=0.7\textwidth,height=0.25\textwidth]{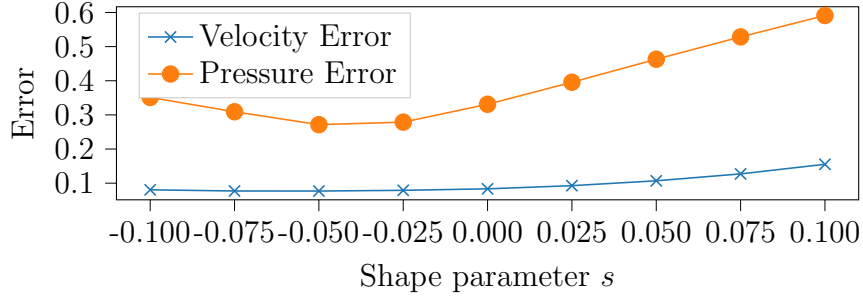}}
	\vfill
	\subfloat[{$R_e=200$}]
	{\includegraphics[width=0.7\textwidth,height=0.25\textwidth]{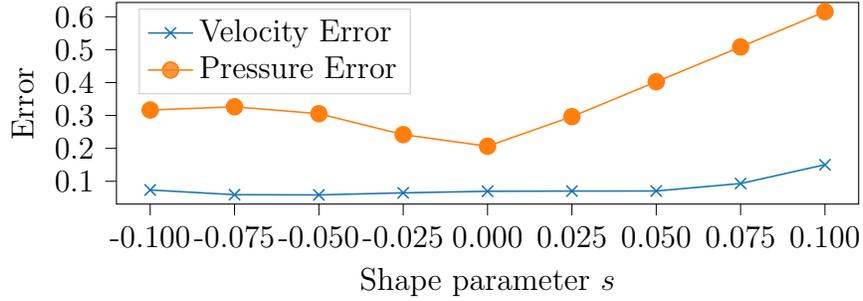}}
	\caption{Relative prediction error of velocity and pressure v.s. the geometry control parameter $s$ ($s>0$ for stenosis and $s<0$ for aneurysm).}
	\label{fig:ParaNSError}
\end{figure}
The relative errors of the predicted velocity and pressure versus geometry parameter $s$ are given in Fig.~\ref{fig:ParaNSError}. We can see that the velocity is accurately predicted with a low relative error, particularly for aneurysm cases ($s<0$). However, the prediction error of pressure in terms of absolute value is relatively large, though the pressure distribution over the entire domain can be captured reasonably well (Figs.~\ref{fig:DetNSContour} and~\ref{fig:DetNSContour_Re200}). For both velocity and pressure, the prediction performance deteriorates as the geometry becomes more stenotic ($s > 0$). 

{The results showed above are from the network trained by seeing all three different families of the input geometries, i.e., straight, stenosis, and aneurysm vessels. Although our label-free PhyGeoNet is not constrained by the availability of labeled data and thus can be freely trained by selecting any input scenarios to leverage the power of DNNs for interpolatory problems, it's still interesting to investigate the ``extrapolation'' performance in case of insufficient training due to computational limitations.}
\begin{figure}[ht]
	\centering
	\includegraphics[width=0.8\textwidth,height=0.4\textwidth]{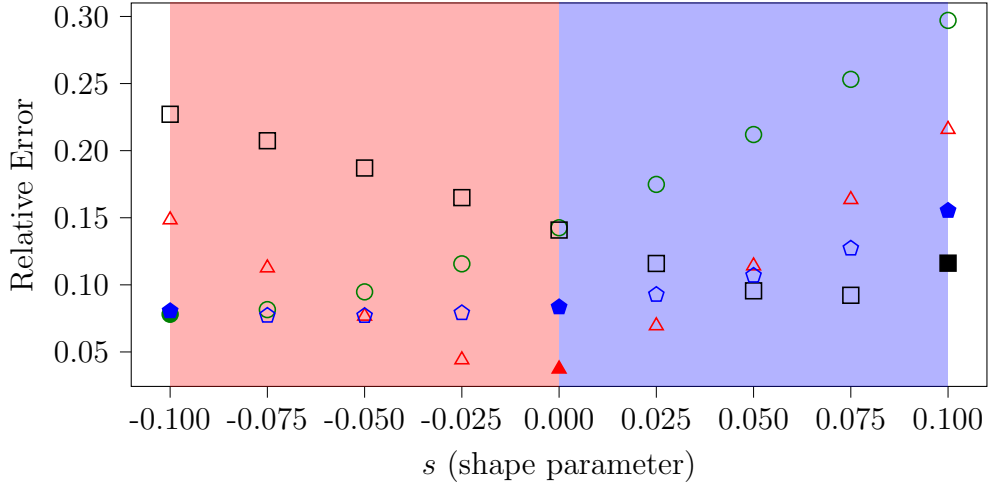}
	\caption{{Comparison of the model performance for four different training scenarios: \textbf{(1)} \ul{training only in the stenosis region} ($s>0$, \ref{plot:stenosis region}) with one training sample (\ref{plot:stenosis training sample}) and eight testing samples (\ref{plot:stenosis testing sample}); \textbf{(2)} \ul{training only in the aneurysm region} ($s<0$, \ref{plot:aneurysm region}) with one training sample (\ref{plot:aneurysm training sample}) and eight testing samples (\ref{plot:aneurysm testing sample}); \textbf{(3)} \ul{training only on the straight vessel} ($s=0$) with one training sample (\ref{plot:straight training sample}) and eight testing samples (\ref{plot:straight testing sample}); \textbf{(4)} \ul{training on both regions} with three training samples (\ref{plot:aneurysm, straight, stenosis, training sample}) and six testing samples (\ref{plot:aneurysm, straight, stenosis, testing sample}).}}
	\label{fig:extrapoaltionlowdim}
\end{figure}
{Here we will compare the prediction performance for four different training scenarios (see Fig.~\ref{fig:extrapoaltionlowdim}): (1) the model is trained only on one input geometry (\ref{plot:stenosis training sample}) in the stenosis region ($s > 0$, \ref{plot:stenosis region}) and evaluated on eight unseen geometries (\ref{plot:stenosis testing sample}) in both stenosis and aneurysm regions; (2) the model is trained only on one input geometry (\ref{plot:stenosis training sample}) in the aneurysm region ($s < 0$, \ref{plot:aneurysm region}) and tested on eight unseen geometries (\ref{plot:stenosis testing sample}) in both stenosis and aneurysm regions; (3) the model is trained on straight vessels (\ref{plot:straight training sample}) and evaluated on eight unseen stenosis and aneurysm vessels (\ref{plot:straight testing sample}); (4) the model is trained on three representative geometries (\ref{plot:aneurysm, straight, stenosis, training sample}) from three families and evaluated on the other six unseen geometries (\ref{plot:aneurysm, straight, stenosis, testing sample}). The comparison results for cases with Reynolds numbers around $Re = 20$ are shown in Fig.~\ref{fig:extrapoaltionlowdim} (The results of the cases with $Re = 200$ are similar and thus are omitted here for brevity.)  As expected, the prediction error increases when the testing points move away from the training input points, similar to what we observed in data-driven cases. It is clear that the model performance is much better for the predictions of interpolation points than that of the extrapolation ones. Since the availability of training labels will not limit our present DNN model, which thus has strong flexibility to select ``important'' input points based on experimental design algorithms~\cite{gao2019bi}, for further optimization of training cost and predictive accuracy.} 

\subsubsection{Poisson equations with spatially-varying source (Case 5)\label{sec:parapoisson}}
\begin{figure}[ht]
	\centering
	\includegraphics[width=0.8\textwidth]{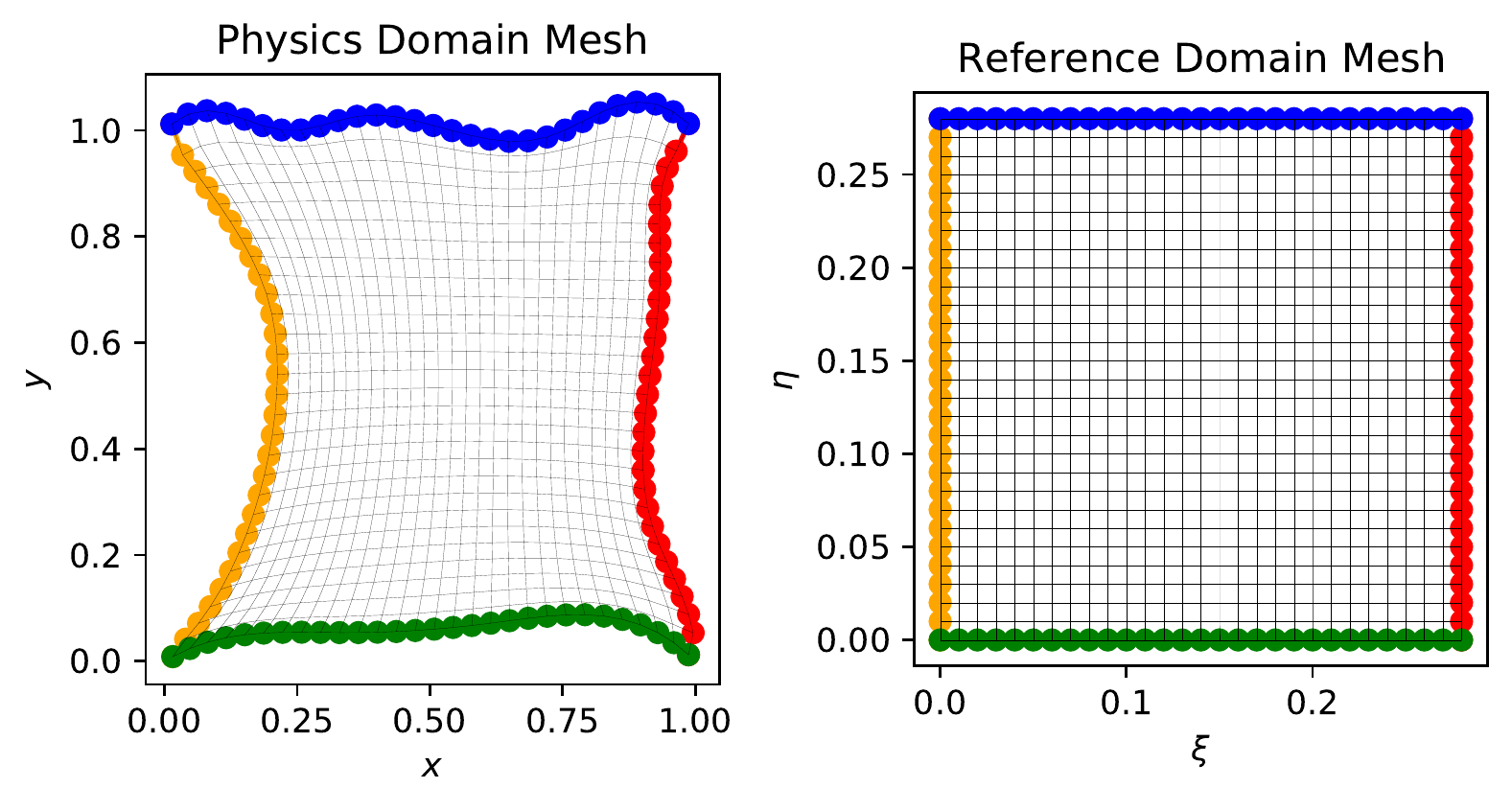}
	\caption{{The physical domain (left) and reference domain (right) for the parametric Poisson equation. The corresponding edges of two domains are indicated by the same color.}}
	\label{fig:MeshPossion}
\end{figure}

\begin{figure}[ht]
	\centering
	\includegraphics[width=0.85\textwidth]{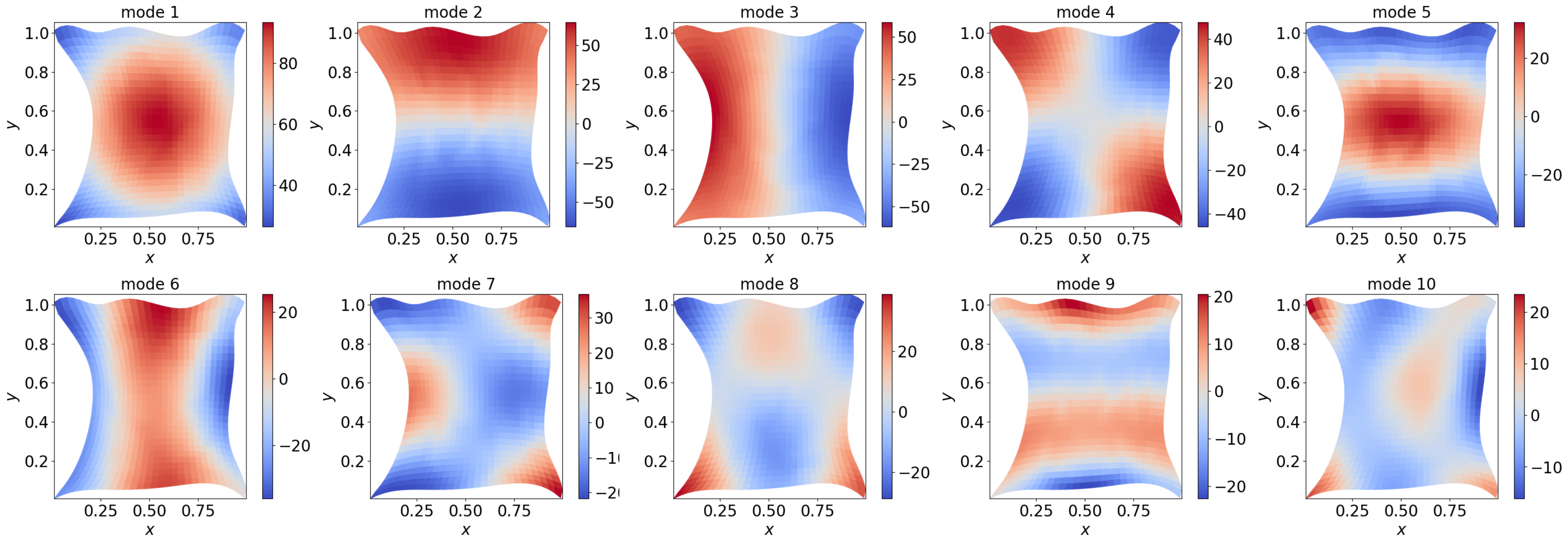}
	\caption{{The first ten K-L mode functions ($\sqrt{\lambda}\phi(\mathbf{x})$) evaluated on the mesh.}}
	\label{fig:KLMODE}
\end{figure}

\begin{figure}[H]
	\centering
	\includegraphics[width=0.98\textwidth]{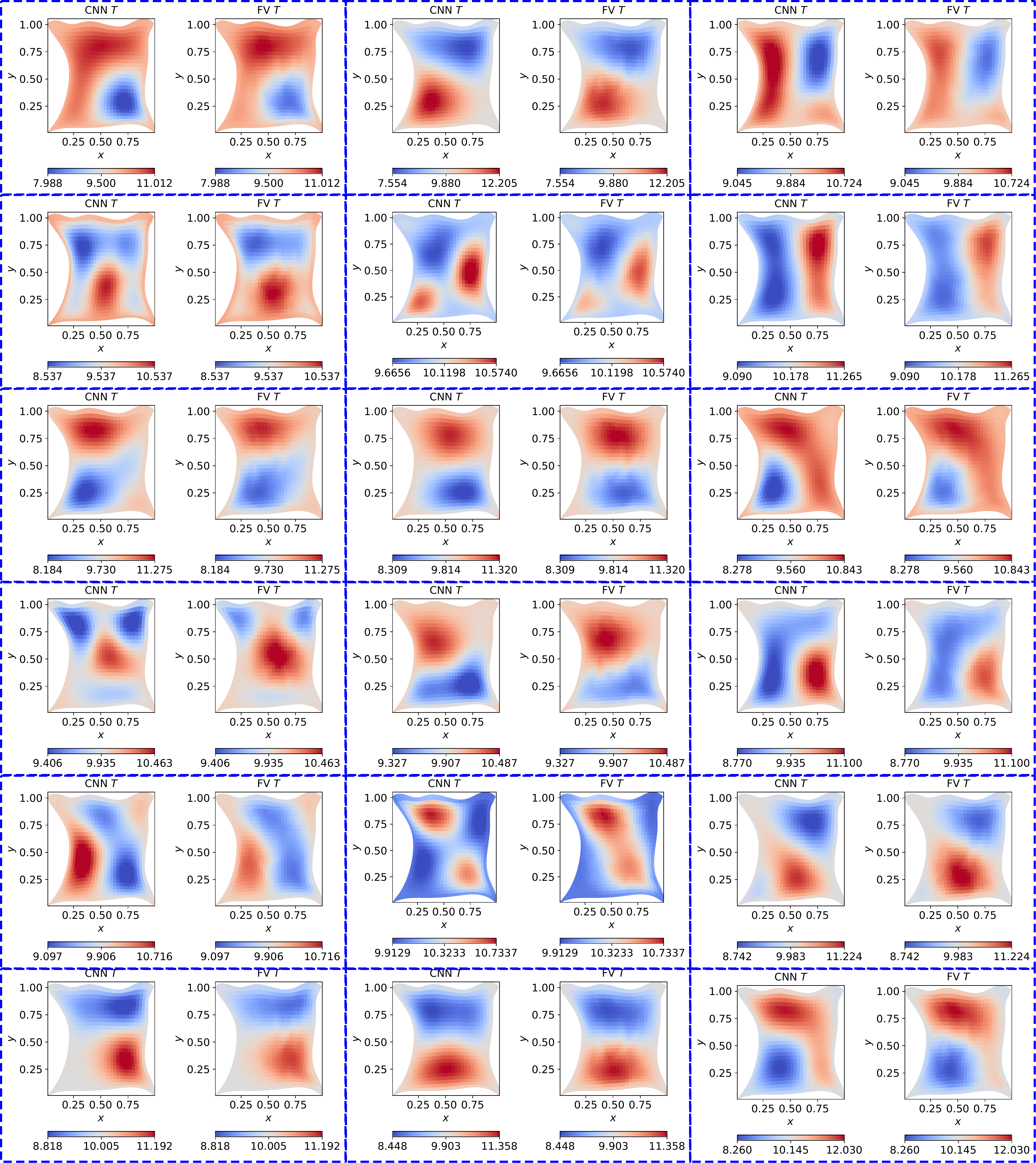}
	\caption{PhyGeoNet solution predictions on a subset of training samples of the spatially-varying source term.}
	\label{fig:PoissonTrainContour}
\end{figure}

\begin{figure}[H]
	\centering
	\includegraphics[width=0.98\textwidth]{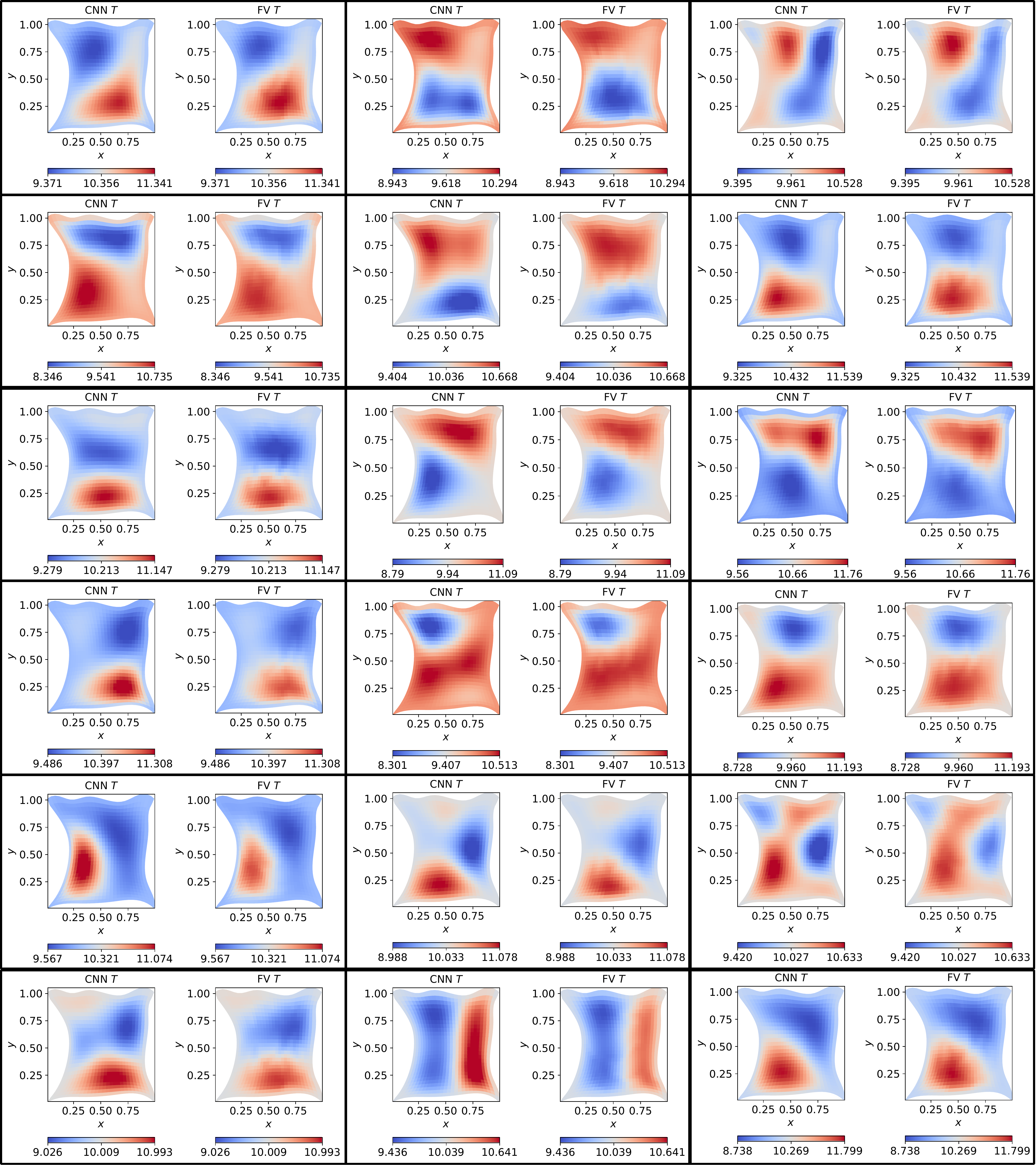}
	\caption{PhyGeoNet solution predictions on a subset of testing samples of the spatially-varying source term.}
	\label{fig:PoissonTestContour}
\end{figure}

{In the last numerical example, we will apply the PhyGeoNet to learn the solutions in a high-dimension parameter space. Consider a Poisson equation (Eqn.~\ref{eqn:posseqn}) with homogeneous Dirichlet boundary conditions ($T_c = 10$) and spatially-varying source term $f(\mathbf{x})$,}
\begin{equation}
	\begin{split}
	\nabla\cdot (\nabla T(\mathbf{x})) + f(\mathbf{x}) =0, \  \mathbf{x}\in\Omega_p,\\
	T(\mathbf{x})=T_c, \ \mathbf{x}\in\partial\Omega_p,
	\end{split}
	\label{eqn:posseqn}
\end{equation}
{where $\Omega_p$ represents the physical domain as shown in Fig.~\ref{fig:MeshPossion}a.
}	 
{The goal here is to learn the parameter-to-solution map $f(\mathbf{x}) \to T(\mathbf{x})$, where the source field has a dimension of  900 ($30\times30$) on the discrete domain. The high-dimensional spatially-varying source term $f(\mathbf{x})$ is modeled as a Gaussian random field,}
\begin{equation}
f(\mathbf{x})\sim\mathcal{GP}(\mathbf{0},K(\mathbf{x},\mathbf{x}')),\quad
K(\mathbf{x},\mathbf{x}')=\sigma_0^2\exp(\frac{(\mathbf{x}-\mathbf{x}')}{2l^2}),
\end{equation}
{where $K(\mathbf{x},\mathbf{x}')$ is the expontential kernel fucntion with the stationary variance of $\sigma_0=100$ and length scale of $l=0.5$. Becuase of the smoothness of the field, the Gaussian process can be expressed in a compact form based on Karhunen-Loeve (K-L) expansion \cite{tipping1999probabilistic},}
\begin{equation}
	f(\mathbf{x})=\sum_{i=1}^{n_k \to \infty}\sqrt{\lambda_i}\phi_i(\mathbf{x})\omega_i,
\end{equation} 
{where $\lambda_i$ and $\phi_i(\mathbf{x})$ are eigenvalues and eigenfunctions of the kernel $K$; $\omega_i$ are an uncorrelated random variables with zero mean and unit variance. The K-L expansion can be truncated to its first ten modes to capture over $99\%$ energy of the original field. Namely, the 900-dimensional spatially-varying source field can be reconstructed based on the coefficients of the first ten K-L modes (Fig.~\ref{fig:KLMODE}), and thus the intrinsic dimension of the parameter space is reduced to ten ( i.e., $\boldsymbol{\omega}=[\omega_1,...,\omega_{10}]\in\mathbb{R}^{10}$). Even so, to capture the solution response surface in a ten-dimensional parameter space is challenging.}
{In our learning framework, the PhyGeoNet takes the input of the source field $f(\mathbf{x})$ and output the solution field($T(\mathbf{x})$). To train the model, we minimize the PDE loss on 256 randomly-sampled input source fields. Once the model is trained, we can not only solve the solutions of the 256 source fields simultaneously, but also build a surrogate model to rapidly predict the solution of any new source field. To test the model, we randomly sample 744 unseen source fields and use the trained PhyGeoNet to predict the corresponding solution fields. The prediction results on a subset of the training and testing samples are shown in Figs.~\ref{fig:PoissonTrainContour} and~\ref{fig:PoissonTestContour}, respectively, compared against the FV-based reference. Overall, the mean relative error for the training samples is $0.0687$ with the variance of $0.0025$. For testing samples, the mean relative error slightly increases to $0.0703$ with the variance of $0.0041$. These results demonstrate the great potential of the PhyGeoNet as the surrogate for high-dimensional problems.
 } 

\section{Discussion}
\label{sec:discussion}

\subsection{Comparison between PhyGeoNet and PINN}
As mentioned in Section~\ref{sec:intro}, the majority of the existing works on PDE-constrained learning have adopted a pointwise formulation using fully-connected neural networks (FC-NN), e.g., physics-informed neural networks (PINN)~\cite{raissi2019physics,sun2020surrogate}. Although the pointwise FC-NN can leverage automatic differentiation to compute derivatives analytically, the training may not be scalable for complex, large-scale problems. The CNN-be\label{key}ased formulations explored in this work has the potential to largely enhance the training efficiency. To demonstrate the merit of the proposed CNN-based PhyGeoNet, quantitative performance comparisons between the PINN and PhyGeoNet are conducted. Specifically, the PINN with the network structure proposed in~\cite{sun2020surrogate} is compared with the PhyGeoNet on solving the steady Navier-Stokes equation defined in Section~\ref{sec:N-S}. Two different scenarios are considered: (1) to compare the training cost of reaching the convergence, and (2) to compare the predictive accuracy with the same training budget. 

\subsubsection{Training cost comparison for convergence}
\begin{figure}[ht]
	\centering
	{\includegraphics[width=0.65\textwidth]{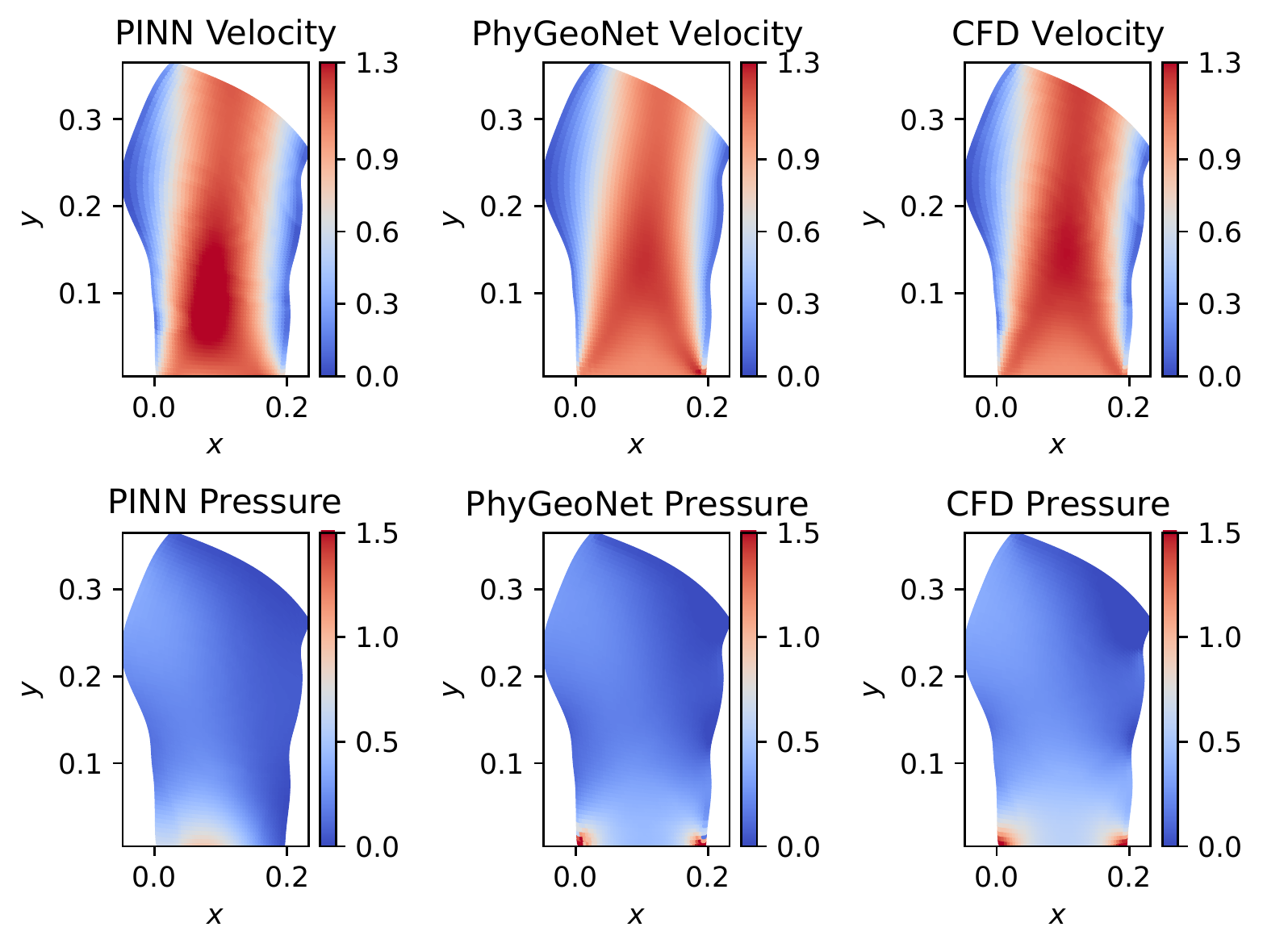}}
	\caption{Velocity (the first row) and pressure (the second row) contours of PINN (left column), PhyGeoNet (middle column), and CFD benchmakrk (right column). Both PINN and PhyGeoNet are fully trained at {$R_e=20$}:}
	\label{fig:ParaNSConvergence}
\end{figure}

We first investigate the training cost and prediction performance of the PINN (FC-NN) and PhyGeoNet (CNN), when both cases are fully converged. The contour comparisons of velocity and pressure solutions are shown in Fig.~\ref{fig:ParaNSConvergence}. Both the PINN and PhyGeoNet can capture the general flow pattern. However, notable discrepancies are observed in the PINN-predicted velocity field, especially in the region where the velocity is developing, while the PhyGeoNet prediction result is more accurate and agree with the CFD benchmark very well. Moreover, the PINN fails to accurately predict the pressure distribution, particularly in the near-inlet zone. In contrast, the PhyGeoNet predicted pressure contour also well agrees with the CFD result.

\begin{table}[H]
	\begin{center}
		\begin{tabular}{ |c|c|c| } 
			\hline
			\multicolumn{1}{|c|}{\diagbox{Information}{ Name}}&\multicolumn{1}{|c|}{PhyGeoNet} &  \multicolumn{1}{|c|}{PINN}\\
			\hline
			\multicolumn{1}{|c|}{Number of iterations}&\multicolumn{1}{|c|}{$1.5\times 10^3$} &  \multicolumn{1}{|c|}{$1.6\times 10^4$} \\
			\hline
			\multicolumn{1}{|c|}{Wall-clock time ($s$)}&\multicolumn{1}{|c|}{$9.12\times 10^2$} &  \multicolumn{1}{|c|}{$9.454\times 10^3$}\\
			\hline
			\multicolumn{1}{|c|}{Velocity relative error}&\multicolumn{1}{|c|}{$0.078$} &  \multicolumn{1}{|c|}{$0.130$}\\
			\hline
			\multicolumn{1}{|c|}{Pressure relative error}&\multicolumn{1}{|c|}{$0.346$} &  \multicolumn{1}{|c|}{$0.632$}\\
			\hline
			\multicolumn{1}{|c|}{Device name}&\multicolumn{2}{|c|}{Nvida GeForce RTX 2080}\\
			\hline
		\end{tabular}
		\caption{Comparison of training costs and prediction performance between PINN and PhyGeoNet, both of which are fully trained.\label{tab:HighlightCost}}
	\end{center}
\end{table}
The comparison of the training costs for both networks being fully trained is listed in Tab.~\ref{tab:HighlightCost}. We found that PINN needs about 10 times more iterations than the PhyGeoNet does to reach the convergence. It takes about 2.5 hours to train the PINN, while it only needs around 15 minutes to train the PhyGeoNet but with higher predictive accuracy. This comparison higyhlights that the PhyGeoNet converges much faster than the PINN does. The main reason is that the convolution operation with shift-invariant filters enables the update of the entire field in each iteration, which is far more efficient than the pointwise training.

\subsubsection{Predictive accuracy comparison with the same training budget}
\begin{figure}[ht]
	\centering
	{\includegraphics[width=0.65\textwidth]{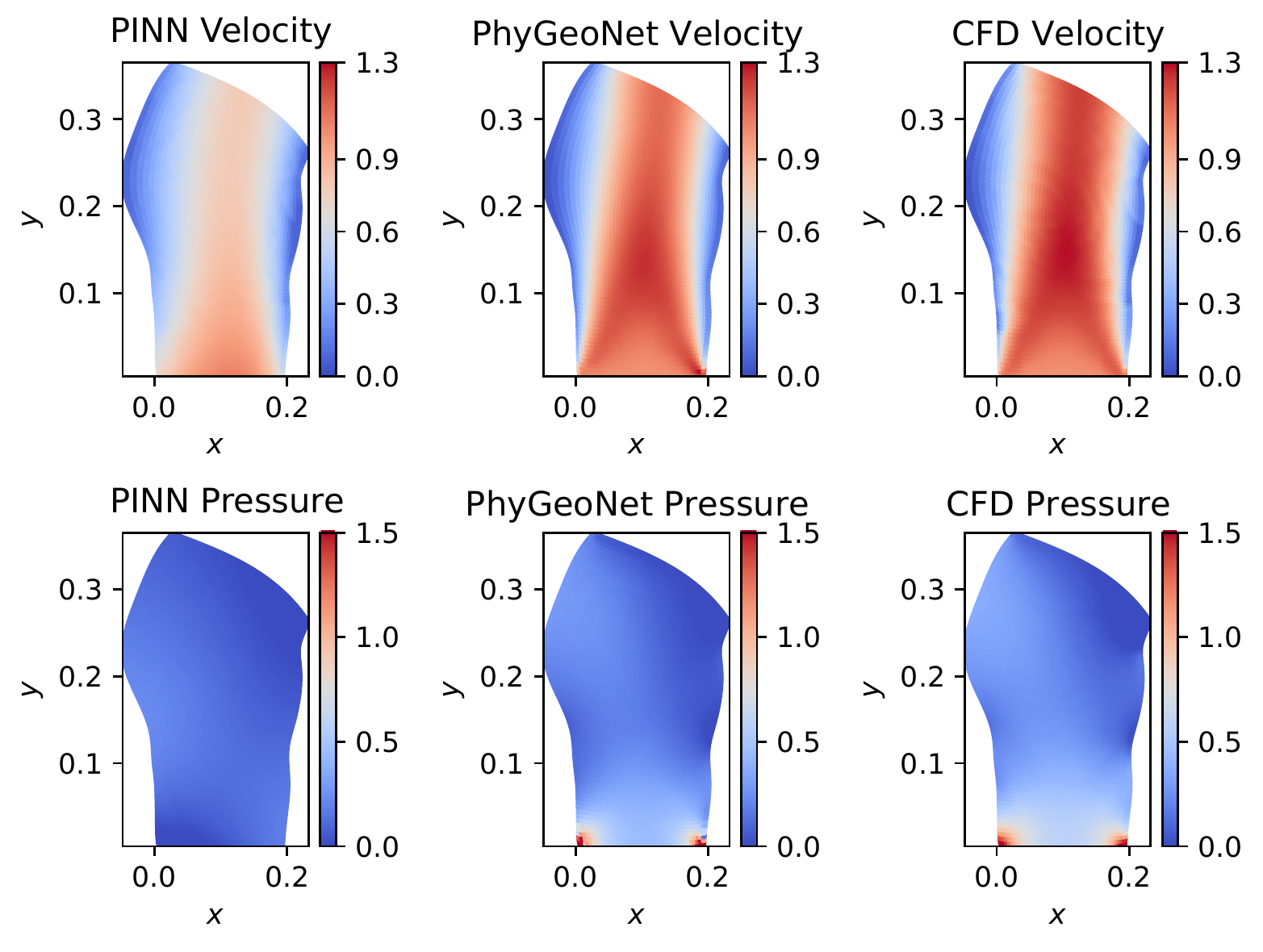}}
	\caption{Velocity (the first row) and pressure (the second row) contours of PINN (left column), PhyGeoNet (middle column), and CFD benchmakrk (right column). Both PINN and PhyGeoNet are trained with the same number of iterations ($\approx 1.6\times10^3$ iterations) at {$R_e=20$}.}
	\label{fig:ParaNSConvergence1}
\end{figure}

To further assess the proposed PhyGeoNet, we conduct another comparison experiment by fixing the total computational budget for training. Namely, both the PhyGeoNet and PINN are trained with a similar amount of iterations, and the training costs (wall time) are thus roughly the same. It can be seen from Fig.~\ref{fig:ParaNSConvergence1} that the PhyGeoNet can accurately predict both velocity and pressure fields after $1.5\times10^3$ training iterations. However, with the same training budget, the PINN's predictions are clearly not accurate and the relative errors of the predicted velocity and pressures are over $30\%$ and $85\%$, respectively, as shown in Tab.~\ref{tab:HighlightAccuracy}.
The comparison study has demonstrated the advantages of the CNN-based PhyGeoNet over the PINN using FC-NN formulations. Since the computational budget is usually limited in practical applications. the PhyGeoNet is preferred due to its fast convergence and higher accuracy. 
\begin{table}[H]
	\begin{center}
		\begin{tabular}{ |c|c|c| } 
			\hline
			\multicolumn{1}{|c|}{\diagbox{Information}{ Name}}&\multicolumn{1}{|c|}{PhyGeoNet} &  \multicolumn{1}{|c|}{PINN}\\
			\hline
			\multicolumn{1}{|c|}{Number of iterations}&\multicolumn{1}{|c|}{$1.5\times 10^3$} &  \multicolumn{1}{|c|}{$1.6\times 10^3$} \\
			\hline
			\multicolumn{1}{|c|}{Wall-clock time ($s$)}&\multicolumn{1}{|c|}{$9.12\times 10^2$} &  \multicolumn{1}{|c|}{$10.34\times 10^2$}\\
			\hline
			\multicolumn{1}{|c|}{Velocity relative error}&\multicolumn{1}{|c|}{$0.078$} &  \multicolumn{1}{|c|}{$0.321$}\\
			\hline
			\multicolumn{1}{|c|}{Pressure relative error}&\multicolumn{1}{|c|}{$0.346$} &  \multicolumn{1}{|c|}{$0.850$}\\
			\hline
			\multicolumn{1}{|c|}{Device name}&\multicolumn{2}{|c|}{Nvida GeForce RTX 2080}\\
			\hline
		\end{tabular}
		\caption{Comparison of predictive accuracy of PINN and PhyGeoNet with the same amount of training budget.\label{tab:HighlightAccuracy}}
	\end{center}
\end{table}

\subsection{Success, limitation, and future perspectives of the current framework}
\begin{figure}[htp]
	\centering
	\includegraphics[width=0.4\textwidth]{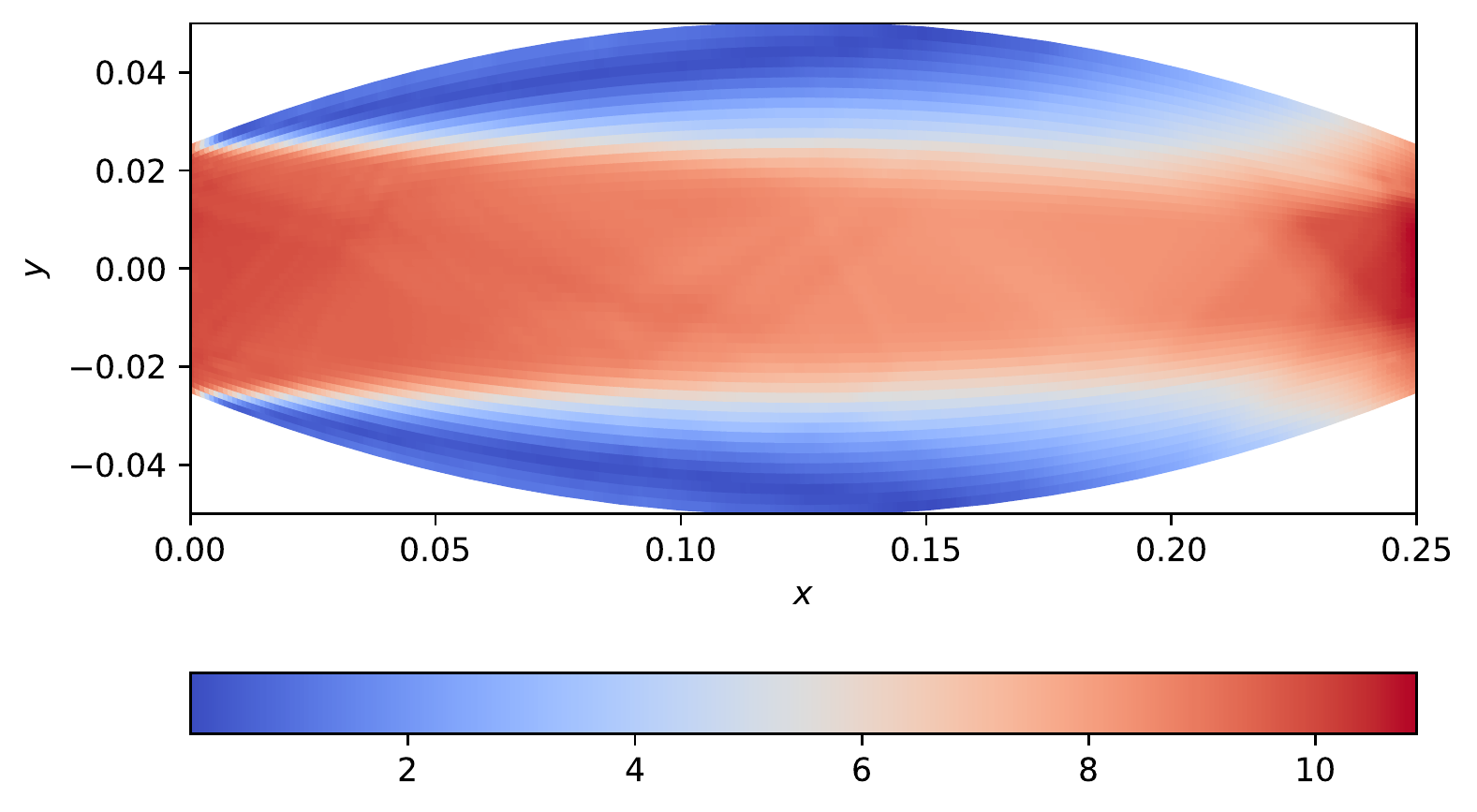}
	\caption{RANS mean flow field of a diverging-converging channel at $Re = 50000$}
	\label{fig:rans}
\end{figure}

\begin{figure}[htp]
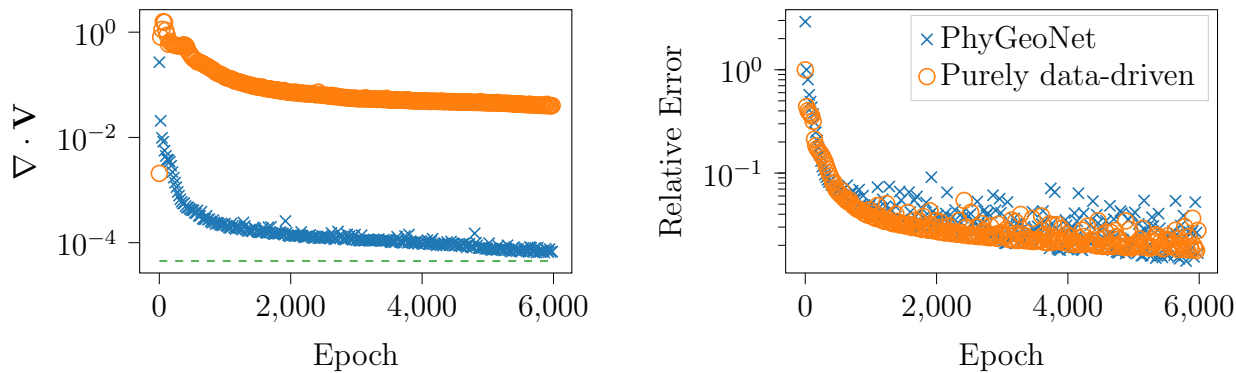

	\centering
	\subfloat[Convergence of mass conservation.]
	{\includegraphics[width=0.48\textwidth,height=0.3\textwidth]{./TurbulentRes.tikz}}
	\hfill
	\subfloat[Convergence history of relative error.]
	{\includegraphics[width=0.48\textwidth,height=0.3\textwidth]{./TurbulentErr.tikz}}
	\caption{Comparison of data-driven CNN solutions with and without the divergence-free constraint for RANS turbulence data on an irregular domain.}
	\label{fig:fittingData}
\end{figure}

{The main contribution of this work lies in the algorithmic development that enables the direct use of classic CNN backbones to solve PDEs on irregular domains. Although the current work focuses on \emph{label-free PDE-driven} learning (i.e., solving non-parametric/parametric PDEs without data), the proposed method can also be applied to enable \emph{data-driven} CNNs with PDE constraints for dealing with problems in irregular domains. For example, the divergence-free condition has been utilized to constrain the data-driven learning for incompressible turbulent flows in recent literature~\cite{mohan2020embedding}, which, however, were limited to image-like regular domains for classic CNNs. The coordinate transformation scheme presented in this work can be applied to impose this constraint on non-rectangular domains, enabling PDE-constrained, data-driven CNN solutions. Here we will use an additional test case to demonstrate this idea. Specifically, we consider a 2D turbulent flow in a non-rectangular domain at $Re=50000$, obtained from the RANS simulation with the $k$-$\epsilon$ turbulence model~\cite{launder1983numerical}. The flow domain and mean velocity field are given in Fig.~\ref{fig:rans}.}
{We use CNN to fit the flow data in a data-driven manner, where the divergence-free constraint is enforced in this irregular domain. As shown in Fig.~\ref{fig:fittingData}, although the training errors reach the same level, the constrained data-driven solution is more physical compared to the purely data-driven one, since the mass conservation is better satisfied. However, a slightly larger oscillation is observed when the training is constrained by the physical laws.} 

{Although showing great promise, the current framework has several limitations, and many technical challenges are still present. \ul{First}, this work focuses on parametrized steady-state PDEs, and the current learning architecture cannot deal with dynamic systems. Nonetheless, the coordinate transformation based DL solutions can also be applied to solve spatiotemporal PDEs with some further developments. For example, we could use the finite difference (e.g., central differencing) to formulate time derivatives in the physics-based loss function and build Long-Short Term Memory (LSTM) network architecture to capture temporal coherence. Recent studies have demonstrated the effectiveness of LSTM for dynamic problems using physics-informed learning~\cite{zhang2020physics}. We also can formulate an auto-regressive network architecture~\cite{geneva_modeling_2020} to learn the temporal dependence, and existing numerical time-stepping schemes, e.g., Euler or Runge-Kutta methods, can be leveraged.} {\ul{Second}, the current approach has challenges to deal with very complex domains (e.g., with more than five $C_0$ continuous edges), since it is difficult to establish a one-to-one mapping for such geometries and the elliptic coordinate transformation cannot be directly applied. To address this issue, we can utilize domain decomposition, which will decompose the complex domain into multiple sub-blocks, and each of them contains four $C_0$ edges. In each sub-block, the current coordinate transformation can be applied, and the continuity between two sub-blocks will be strictly imposed using CNN padding.} {\ul{Third}, we acknowledge that our current framework has not been applied to solve the Navier-Stokes equations in turbulence regimes without labels. To achieve this goal, grant technical challenges are present and significant developments are expected: (1) if directly resolving the Navier-Stokes equations at turbulence levels, a very high-resolution mesh is required to resolve the Kolmogorov scales, which would cause computational and memory issues in the current formulation; (2) if modeling instead of resolving sub-scale turbulence structures, appropriate closure or subgrid-scale models need to be built into the networks, and corresponding model-form uncertainty should be considered.} {We will explore these directions in the future work and continue to fill the current gaps for more complex and realistic applications.}

\section{Conclusion}
\label{sec:conclusion}
This paper presented a novel method of physics-informed CNN (PhyGeoNet) for solving parametric PDEs on irregular domains without any labeled data. An elliptic mapping is introduced to formulate the transformation between the irregular physical domain and regular reference domain, enabling the direct use of powerful classic CNN backbones for non-rectangular geometries and non-uniform grids. Since the elliptic mapping is obtained numerically, the proposed method can be applied to complex geometries that are not parameterizable. The effectiveness and merit of the proposed PhyGeoNet have been demonstrated by solving a number of PDE systems on irregular domains, including nonparametric/parametric heat equations and Navier-Stokes equations. Moreover, the PhyGeoNet was compared with the state-of-the-art PINN in terms of accuracy and efficiency. The comparison results have shown that the convergence speed of the PhyGeoNet is more than an order of magnitude faster than that of the PINN (with FC-NN formulation), and the accuracy of the PhyGeoNet is much higher if the total training budget is fixed to be the same. These advantages highlight the potential of PhyGeoNet on solving more complex problems that are scalable.

\section*{Acknowledgment}
The authors would like to acknowledge the funds from National Science Foundation (NSF contract CMMI-1934300) and startup funds from the College of Engineering at University of Notre Dame in supporting this study. We also gratefully acknowledge the discussion with Dr. Matthew J. Zahr in the early stage of this research. {The authors finally thank the anonymous reviewers for their insightful comments and suggestions to improve the quality of this paper.}


\appendix
\section{Proof of forward elliptic transformation \label{sec:proofINVMap}}
\begin{proof}
The derivatives of reference coordinates with respect to physical coordinates can be expressed as~\cite{anderson1995computational},
\begin{linenomath*}
	\begin{subequations}
		\label{eqn:matrics}
		\begin{alignat}{2}
			\frac{\partial\xi}{\partial x}&=\frac{1}{J}\frac{\partial y}{\partial \eta},\\
			\frac{\partial\eta}{\partial x}&=-\frac{1}{J}\frac{\partial y}{\partial \xi},\\
			\frac{\partial\xi}{\partial y}&=-\frac{1}{J}\frac{\partial x}{\partial \eta},\\
			\frac{\partial \eta}{\partial y}&=\frac{1}{J}\frac{\partial x}{\partial \xi},	
		\end{alignat}
	\end{subequations}
\end{linenomath*}
where $J=\frac{\partial x}{\partial\xi}\frac{\partial y}{\partial\eta}-\frac{\partial x}{\partial\eta}\frac{\partial y}{\partial\xi}\neq0$ is the determinant of the Jacobian matrix. 
 
Based on Eq.~\ref{eqn:Du}, derivatives of the forward map $\mathcal{G}$ can be calculated. Substitute Eq.~\ref{eqn:matrics}a into Eq.~\ref{eqn:Du}a, we have
\begin{equation}
\label{eqn:d2xidx2}
\begin{split}
\frac{\partial^2\xi}{\partial x^2}=&\frac{1}{J^3}\Big[-\frac{\partial^2x}{\partial\eta^2}\left(\frac{\partial y}{\partial\xi}\right)^2\frac{\partial y}{\partial\eta} + \frac{\partial^2 y}{\partial\eta^2}\frac{\partial x}{\partial\eta}\left(\frac{\partial y}{\partial\xi}\right)^2 +2\frac{\partial^2x}{\partial\xi\partial\eta}\frac{\partial y}{\partial\xi}\left(\frac{\partial y}{\partial\eta}\right)^2 - \\
&2\frac{\partial^2 y}{\partial\xi\partial\eta}\frac{\partial x}{\partial\eta}\frac{\partial y}{\partial\xi}\frac{\partial y}{\partial\eta}-\frac{\partial^2x}{\partial\xi^2}\left(\frac{\partial y}{\partial\eta}\right)^3 + \frac{\partial^2 y}{\partial\xi^2}\frac{\partial x}{\partial\eta}\left(\frac{\partial y}{\partial\eta}\right)^2\Big]
\end{split}
\end{equation}
Substitute Eq.~\ref{eqn:matrics}c into Eq.~\ref{eqn:Du}b, we can get
\begin{equation}
\label{eqn:d2xidy2}
\begin{split}
\frac{\partial^2\xi}{\partial y^2}=&\frac{1}{J^3}\Big[\frac{\partial^2y}{\partial\eta^2}\left(\frac{\partial x}{\partial\xi}\right)^2\frac{\partial x}{\partial\eta} - \frac{\partial^2 x}{\partial\eta^2}\frac{\partial y}{\partial\eta}\left(\frac{\partial x}{\partial\xi}\right)^2 -2\frac{\partial^2y}{\partial\xi\partial\eta}\frac{\partial x}{\partial\xi}\left(\frac{\partial x}{\partial\eta}\right)^2 + \\
&2\frac{\partial^2 x}{\partial\xi\partial\eta}\frac{\partial y}{\partial\eta}\frac{\partial x}{\partial\xi}\frac{\partial x}{\partial\eta}+\frac{\partial^2y}{\partial\xi^2}\left(\frac{\partial x}{\partial\eta}\right)^3 - \frac{\partial^2 x}{\partial\xi^2}\frac{\partial y}{\partial\eta}\left(\frac{\partial x}{\partial\eta}\right)^2\Big]
\end{split}
\end{equation}
Substitute Eq.~\ref{eqn:d2xidx2} and Eq.~\ref{eqn:d2xidy2} into Eq.~\ref{eqn:GInv} and then multiply $J^3$ on both side, we get
\begin{equation}
\label{eqn:invDeltaXi}
\begin{split}
&-\frac{\partial^2x}{\partial\eta^2}\left(\frac{\partial y}{\partial\xi}\right)^2\frac{\partial y}{\partial\eta} + \frac{\partial^2 y}{\partial\eta^2}\frac{\partial x}{\partial\eta}\left(\frac{\partial y}{\partial\xi}\right)^2 +2\frac{\partial^2x}{\partial\xi\partial\eta}\frac{\partial y}{\partial\xi}\left(\frac{\partial y}{\partial\eta}\right)^2 - \\
&2\frac{\partial^2 y}{\partial\xi\partial\eta}\frac{\partial x}{\partial\eta}\frac{\partial y}{\partial\xi}\frac{\partial y}{\partial\eta}-\frac{\partial^2x}{\partial\xi^2}\left(\frac{\partial y}{\partial\eta}\right)^3 + \frac{\partial^2 y}{\partial\xi^2}\frac{\partial x}{\partial\eta}\left(\frac{\partial y}{\partial\eta}\right)^2 +\\
&\frac{\partial^2y}{\partial\eta^2}\left(\frac{\partial x}{\partial\xi}\right)^2\frac{\partial x}{\partial\eta} - \frac{\partial^2 x}{\partial\eta^2}\frac{\partial y}{\partial\eta}\left(\frac{\partial x}{\partial\xi}\right)^2 -2\frac{\partial^2y}{\partial\xi\partial\eta}\frac{\partial x}{\partial\xi}\left(\frac{\partial x}{\partial\eta}\right)^2 + \\
&2\frac{\partial^2 x}{\partial\xi\partial\eta}\frac{\partial y}{\partial\eta}\frac{\partial x}{\partial\xi}\frac{\partial x}{\partial\eta}+\frac{\partial^2y}{\partial\xi^2}\left(\frac{\partial x}{\partial\eta}\right)^3 - \frac{\partial^2 x}{\partial\xi^2}\frac{\partial y}{\partial\eta}\left(\frac{\partial x}{\partial\eta}\right)^2=0
\end{split}
\end{equation} 
Then substitute Eq.~\ref{eqn:matrics}b into Eq.~\ref{eqn:Du}a, we get
\begin{equation}
\label{eqn:d2etadx2}
\begin{split}
\frac{\partial^2\eta}{\partial x^2}=&\frac{1}{J^3}\Big[\frac{\partial^2x}{\partial\eta^2}\left(\frac{\partial y}{\partial\xi}\right)^3
-2\frac{\partial^2 x}{\partial\eta\partial\xi}\left(\frac{\partial y}{\partial\xi}\right)^2\frac{\partial y}{\partial\eta} 
-\frac{\partial^2y}{\partial\eta^2}\frac{\partial x}{\partial\xi}\left(\frac{\partial y}{\partial\xi}\right)^2 \\
&+\frac{\partial^2 x}{\partial\xi^2}\frac{\partial y}{\partial\xi}\left(\frac{\partial y}{\partial\eta}\right)^2
+2\frac{\partial^2y}{\partial\xi\partial\eta}\frac{\partial x}{\partial\xi}\frac{\partial y}{\partial\xi}\frac{\partial y}{\partial\eta} 
-\frac{\partial^2 y}{\partial\xi^2}\frac{\partial x}{\partial\xi}\left(\frac{\partial y}{\partial\eta}\right)^2\Big].
\end{split}
\end{equation}
Substitute Eq.~\ref{eqn:matrics}d into Eq.~\ref{eqn:Du}b, we get
\begin{equation}
\label{eqn:d2etady2}
\begin{split}
\frac{\partial^2\eta}{\partial y^2}=&\frac{1}{J^3}\Big[-\frac{\partial^2y}{\partial\eta^2}\left(\frac{\partial x}{\partial\xi}\right)^3
+2\frac{\partial^2 y}{\partial\eta\partial\xi}\left(\frac{\partial x}{\partial\xi}\right)^2\frac{\partial x}{\partial\eta} 
+\frac{\partial^2x}{\partial\eta^2}\frac{\partial y}{\partial\xi}\left(\frac{\partial x}{\partial\xi}\right)^2 \\
&-\frac{\partial^2 y}{\partial\xi^2}\frac{\partial x}{\partial\xi}\left(\frac{\partial x}{\partial\eta}\right)^2
-2\frac{\partial^2x}{\partial\xi\partial\eta}\frac{\partial y}{\partial\xi}\frac{\partial x}{\partial\xi}\frac{\partial x}{\partial\eta} 
+\frac{\partial^2 x}{\partial\xi^2}\frac{\partial y}{\partial\xi}\left(\frac{\partial x}{\partial\eta}\right)^2\Big].
\end{split}
\end{equation}
Substitute Eq.~\ref{eqn:d2etadx2} and Eq.~\ref{eqn:d2etady2} into Eq.~\ref{eqn:GInv} and then multiply $J^3$ on both side, we get
\begin{equation}
\label{eqn:invDeltaEta}
\begin{split}
&\frac{\partial^2x}{\partial\eta^2}\left(\frac{\partial y}{\partial\xi}\right)^3
-2\frac{\partial^2 x}{\partial\eta\partial\xi}\left(\frac{\partial y}{\partial\xi}\right)^2\frac{\partial y}{\partial\eta} 
-\frac{\partial^2y}{\partial\eta^2}\frac{\partial x}{\partial\xi}\left(\frac{\partial y}{\partial\xi}\right)^2 \\
&+\frac{\partial^2 x}{\partial\xi^2}\frac{\partial y}{\partial\xi}\left(\frac{\partial y}{\partial\eta}\right)^2
+2\frac{\partial^2y}{\partial\xi\partial\eta}\frac{\partial x}{\partial\xi}\frac{\partial y}{\partial\xi}\frac{\partial y}{\partial\eta} 
-\frac{\partial^2 y}{\partial\xi^2}\frac{\partial x}{\partial\xi}\left(\frac{\partial y}{\partial\eta}\right)^2+\\
&-\frac{\partial^2y}{\partial\eta^2}\left(\frac{\partial x}{\partial\xi}\right)^3
+2\frac{\partial^2 y}{\partial\eta\partial\xi}\left(\frac{\partial x}{\partial\xi}\right)^2\frac{\partial x}{\partial\eta} 
+\frac{\partial^2x}{\partial\eta^2}\frac{\partial y}{\partial\xi}\left(\frac{\partial x}{\partial\xi}\right)^2 \\
&-\frac{\partial^2 y}{\partial\xi^2}\frac{\partial x}{\partial\xi}\left(\frac{\partial x}{\partial\eta}\right)^2
-2\frac{\partial^2x}{\partial\xi\partial\eta}\frac{\partial y}{\partial\xi}\frac{\partial x}{\partial\xi}\frac{\partial x}{\partial\eta} 
+\frac{\partial^2 x}{\partial\xi^2}\frac{\partial y}{\partial\xi}\left(\frac{\partial x}{\partial\eta}\right)^2=0
\end{split}
\end{equation}
Multiply Eq.~\ref{eqn:invDeltaEta} by $\frac{\partial y}{\partial \eta}$, multiply Eq.~\ref{eqn:invDeltaXi} by $\frac{\partial y}{\partial \xi}$ and sum together we can finally prove Eq.~\ref{eqn:G}b with the fact $J\neq0$,
\begin{equation}
J\left(\alpha\frac{\partial^2y}{\partial\xi^2}-2\beta\frac{\partial^2y}{\partial\xi\partial\eta}+\gamma\frac{\partial^2y}{\partial\eta^2}\right)=0.
\end{equation} 
Multiply Eq.~\ref{eqn:invDeltaEta} by $\frac{\partial x}{\partial \eta}$, multiply Eq.~\ref{eqn:invDeltaXi} by $\frac{\partial x}{\partial \xi}$ and sum together we can finally prove Eq.~\ref{eqn:G}a with the fact $J\neq0$,
\begin{equation}
J\left(\alpha\frac{\partial^2x}{\partial\xi^2}-2\beta\frac{\partial^2x}{\partial\xi\partial\eta}+\gamma\frac{\partial^2x}{\partial\eta^2}\right)=0.
\end{equation}
\end{proof}
Note that, $J$ is not fixed at the stage of building mappings, we can calculate its partial derivatives via Eq.~\ref{eqn:Du}a and Eq.~\ref{eqn:Du}b to get its derivatives,
\begin{subequations}
\begin{equation}
\frac{\partial J}{\partial \xi}=\frac{\partial^2x}{\partial \xi^2}\frac{\partial y}{\partial\eta}+
\frac{\partial x}{\partial\xi}\frac{\partial^2y}{\partial\eta\partial\xi}-
\frac{\partial^2y}{\partial\xi^2}\frac{\partial x}{\partial\eta}-
\frac{\partial y}{\partial\xi}\frac{\partial^2x}{\partial\eta\partial\xi},	     
\end{equation}
\begin{equation}
\frac{\partial J}{\partial \eta}=\frac{\partial^2x}{\partial\xi\partial\eta}\frac{\partial y}{\partial\eta}+
\frac{\partial x}{\partial\xi}\frac{\partial^2y}{\partial\eta^2}-
\frac{\partial^2y}{\partial\xi\partial\eta}\frac{\partial x}{\partial\eta}-
\frac{\partial y}{\partial\xi}\frac{\partial^2x}{\partial\eta^2}.	     
\end{equation} 
\end{subequations}
However, after the mapping is built up, $J$ remains as a constant non-zero point-wise value.

\section{Convolution filter for derivatives on reference domain\label{sec:filterDetail}}
\begin{figure}[H]
	\centering
	{\includegraphics[width=0.8\textwidth]{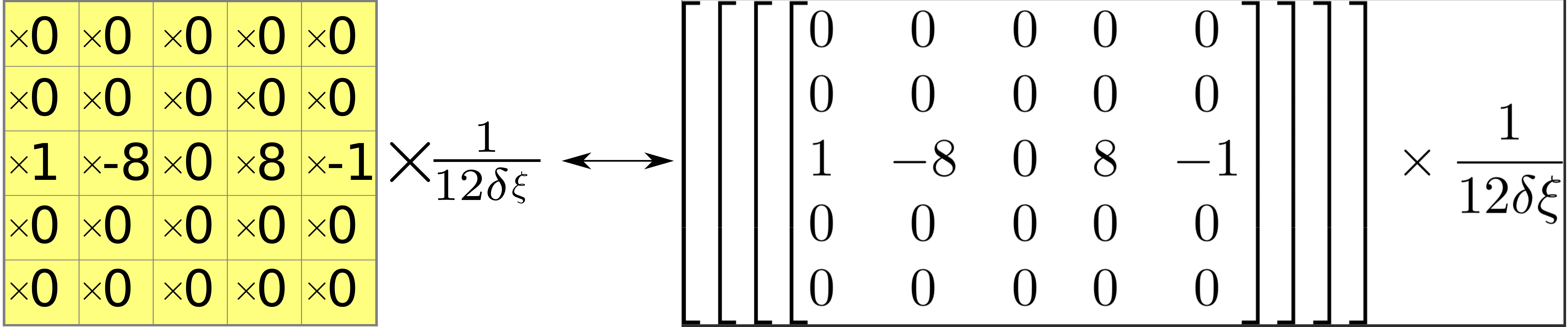}}
	\caption{Finite difference based convolution filter for the differential operator: $\frac{\partial }{\partial \xi}$}
	\label{fig:FilterDemo}
\end{figure}

For internal nodes on the reference domain, the first derivatives are approximated by $4^{\text{th}}$ order central differences,
\begin{linenomath*}
	\begin{subequations}
		\begin{alignat}{2}
	\frac{\partial u}{\partial \xi}\approx\frac{-u_{\xi+2\delta \xi,\eta}+8u_{\xi+\delta \xi,\eta}-8u_{\xi-\delta \xi,\eta}+u_{\xi-2\delta \xi,\eta}}{12\delta \xi} + O((\delta \xi)^4),\\
	\frac{\partial u}{\partial \eta}\approx\frac{-u_{\xi,\eta+2\delta \eta}+8u_{\xi,\eta+\delta \eta}-8u_{\xi,\eta-\delta \eta}+u_{\xi,\eta-2\delta \eta}}{12\delta \eta}+ O((\delta \eta)^4).
		\end{alignat}
	\end{subequations}
\end{linenomath*}
which can be expressed by an convolution filter as shown in Fig.~\ref{fig:FilterDemo}.
For nodes near and on the boundary, one-sided differences (i.e., upwind/downwind) are applied, and thus the artifacts due to padding (i.e., ghost cells) can be avoided. For example, third-order one-sided finite differences on the lower and left boundaries are given by 
\begin{linenomath*}
	\begin{subequations}
		\label{eq:lower_1side}
		\begin{alignat}{2}
		\frac{\partial u}{\partial \xi}\approx\frac{-11u_{\xi,\eta}+18u_{\xi+\delta \xi,\eta}-9u_{\xi+2\delta\xi,\eta}+2u_{\xi+3\delta \xi,\eta}}{6\delta \xi}+O((\delta \xi)^3),\\
		\frac{\partial u}{\partial \eta}\approx\frac{-11u_{\xi,\eta}+18u_{\xi,\eta+\delta \eta}-9u_{\xi,\eta+2\delta \eta}+2u_{\xi,\eta+3\delta\eta}}{12\delta \eta}+O((\delta \eta)^3).
		\end{alignat}
	\end{subequations}
\end{linenomath*}

\section{Compuational hyperparameter setting for PhyGeoNet\label{Appendix:HPS}}
\begin{table}[H]
	\centering
	\small
	\begin{tabular}{|c|c|c|c|c|c|}
		\hline
		\multicolumn{1}{|c|}{\diagbox{Hyperparameters}{Case name}}&
		\multicolumn{1}{|c|}{Case1}&
		\multicolumn{1}{|c|}{Case2}&
		\multicolumn{1}{|c|}{Case3}&
		\multicolumn{1}{|c|}{Case4}&
		\multicolumn{1}{|c|}{{Case5}}\\
		\hline
		\multicolumn{1}{|c|}{Number of iterations for training}&
		\multicolumn{1}{|c|}{1300}&
		\multicolumn{1}{|c|}{15000}&
		\multicolumn{1}{|c|}{1000}&
		\multicolumn{1}{|c|}{20000}&
		\multicolumn{1}{|c|}{{87252}}\\
		\hline
		\multicolumn{1}{|c|}{Number of training parameters}&
		\multicolumn{1}{|c|}{1}&
		\multicolumn{1}{|c|}{1}&
		\multicolumn{1}{|c|}{2}&
		\multicolumn{1}{|c|}{3}&
		\multicolumn{1}{|c|}{{256}}\\
		\hline
		\multicolumn{1}{|c|}{Number of testing parameters}&
		\multicolumn{1}{|c|}{1}&
		\multicolumn{1}{|c|}{1}&
		\multicolumn{1}{|c|}{7}&
		\multicolumn{1}{|c|}{9}&
		\multicolumn{1}{|c|}{{744}}\\
		\hline
		\multicolumn{1}{|c|}{Batch size}&
		\multicolumn{1}{|c|}{1}&
		\multicolumn{1}{|c|}{1}&
		\multicolumn{1}{|c|}{2}&
		\multicolumn{1}{|c|}{3}&
		\multicolumn{1}{|c|}{{32}}\\
		\hline
		\multicolumn{1}{|c|}{Learning Rate}&
		\multicolumn{5}{|c|}{0.001}\\
		\hline
	\end{tabular}
	\caption{Hyperparameters setting for PhyGeoNet in each numerical experiment.}
	\label{tab:ParamterSummary}
\end{table} 
\section{Convergence History of PhyGeoNet Training \label{Appendix:ConvergernceHistory}}
\begin{figure}[H]
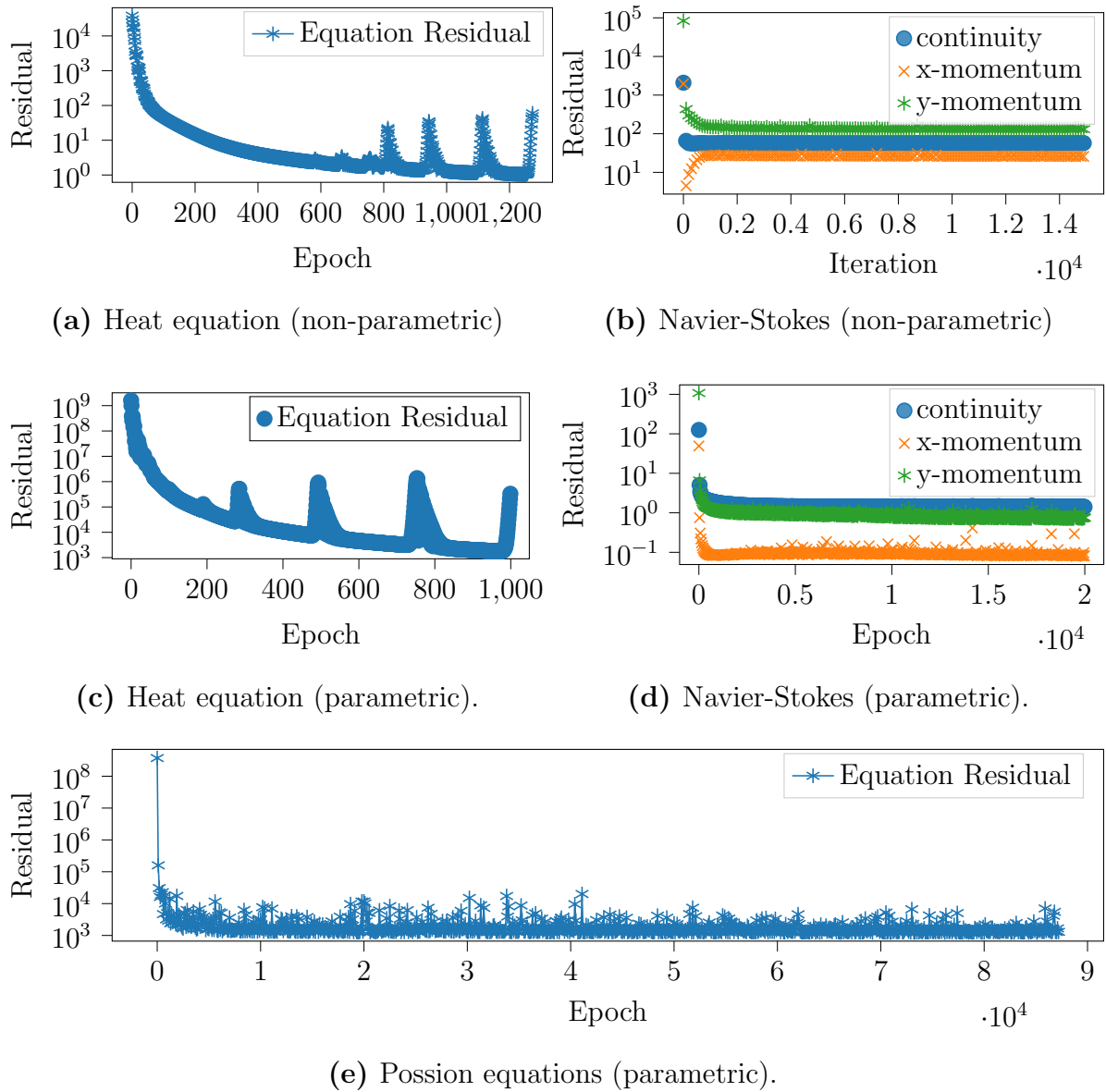

	\centering
	\subfloat[Heat equation (non-parametric)]
	{\includegraphics[width=0.48\textwidth,height=0.24\textwidth]{./DetHeat_convergence.tikz}}
	\subfloat[Navier-Stokes (non-parametric)]
	{\includegraphics[width=0.48\textwidth,height=0.25\textwidth]{./DetNS_Res15000.tikz}}
	\vfill
	\subfloat[Heat equation (parametric).]
	{\includegraphics[width=0.48\textwidth,height=0.24\textwidth]{./ParaHeat_Res1000.tikz}}
	\subfloat[Navier-Stokes (parametric).]
	{\includegraphics[width=0.48\textwidth,height=0.25\textwidth]{./ParaNS_Res20000.tikz}}
	\vfill
	\subfloat[Possion equations (parametric).]
	{\includegraphics[width=0.96\textwidth,height=0.25\textwidth]{./high_dimension_case_convergence.tikz}}
	\caption{Convergence histories of the PhyGeoNet Training.\label{fig:trainingloss}}
\end{figure}

%
%


\end{document}